\documentclass{cernrep}
\begin{document}
\title{Topics in Higgs Physics}
\author{John Ellis\thanks
                 {Also Theoretical Physics Department, CERN, CH 1211 Geneva 23, Switzerland}}
\institute{Theoretical Particle Physics \& Cosmology Group,
Department of Physics,
King's College London,
Strand, London WC2R 2LS, United Kingdom}
\maketitle

\begin{abstract}
These lecture notes review the theoretical background to the Higgs boson, provide an introduction
to its phenomenology, and describe the experimental tests that lead us to think that ``beyond any
reasonable doubt, it is a Higgs boson". Motivations for expecting new physics beyond the
Standard Model are recalled, and the Standard Model effective field theory is advocated as
a tool to help search for it. The phenomenology of $N = 1$ and $N = 2$ supersymmetric Higgs bosons is
reviewed, and the prospects for possible future Higgs factories are previewed. \\
~~\\
KCL-PH-TH/2017-09, CERN-TH/2017-039
 
\end{abstract}


\vspace{0.5cm}

\section{Background to the Higgs Discovery}

\subsection{Historical introduction}

The fundamental equations of physics have a high degree of symmetry -
think the rotation and translation symmetry of Newton's equations,
the gauge invariance of Maxwell's equations for electrodynamics,
the boost symmetry of special relativity, or the the general coordinate invariance
of general relativity - and these theories are generally considered to be
very beautiful. However, the solutions to these equations often conceal
these symmetries, and they appear asymmetric - people are not spherical,
for example. Sometimes these asymmetric solutions may even appear more
beautiful than symmetric solutions - the image of the Mona Lisa, for example, would not
be so interesting if it were symmetric. Regardless of these 
aesthetic considerations, the rich variety of physical phenomena clearly
requires the potential to break symmetries.

However, breaking a symmetry must be done with care - for example, the
gauge invariance of electrodynamics guarantees the renormalisability
of quantum electrodynamics (QED) and hence its calculability. The trick
in formulating theories with `broken' symmetry is often to hide the symmetry
so that it is not manifest, while maintaining it at a fundamental level and
thereby preserving its attractive features such as renormalizability.
This can be done by postulating a lowest-energy (`vacuum') state of the
symmetric equations that does not possess the full symmetry of the
underlying equations - an idea known as spontaneously-broken or
`hidden' symmetry.

This idea originated in condensed-matter physics - an early example
being the superfluidity that plays an essential r\^ole in the LHC magnet
system. In this case, the spontaneously-broken symmetry is global,
i.e., the symmetry transformations are independent of the spatial
location within the superfluid. This type of hidden symmetry was
introduced into particle physics by Yoichiro Nambu~\cite{Nambu}, who used it to understand
the (relatively) low mass and the low-energy dynamics of the pion.
According to his theory, the underlying symmetry (in this case chiral 
symmetry) is not manifest, but is reflected in the couplings of the pion, 
which would have no mass if the up and down quarks were strictly 
massless. Jeffrey Goldstone subsequently published a simple elementary
field-theoretical model of this phenomenon~\cite{Goldstone} and he, 
Abdus Salam and Steven Weinberg~\cite{GSW} subsequently proved
rigorously that in a relativistic theory every spontaneously-broken 
global symmetry would be reincarnated in a massless particle with
specific couplings, called a Nambu-Goldstone boson.

All well and good, but there are interesting cases where the spontaneously-broken
symmetry is local, as in a gauge theory - the prime example being
the superconductivity that also plays an essential r\^ole in the LHC magnet
system. The theory of spontaneous gauge symmetry breaking 
in this non-relativistic situation was first developed by Philip Anderson~\cite{Anderson}
and Nambu~\cite{Nambu2}. According to their theory, inside a superconductor the
(externally massless) photon acquires a medium-dependent mass by
`eating' Cooper pairs of electrons in the lowest-energy (`vacuum') 
state inside the medium. Anderson also conjectured that a similar
phenomenon might be possible in a relativistic theory, but did not 
develop this idea. Indeed, Walter Gilbert~\cite{Gilbert} argued that this would {\it not}
be possible, because spontaneous symmetry breaking seemed to
require the presence of a vector breaking Lorentz symmetry
explicitly.

However, in 1964 this argument was circumvented in papers by Fran\c{c}ois Englert 
and Robert Brout~\cite{EB}, and by Peter Higgs~\cite{Higgs1,Higgs2}. The Englert-Brout paper was received by
the journal where it was published on June 26th, 1964, and a first
paper by Higgs was received on July 27th, 1964. Unaware of the paper
by Englert and Brout, he
pointed out a potential loophole in the Gilbert argument, and in
a second paper he constructed an explicit example. Curiously, 
whereas the first paper was accepted quickly by the journal Physics
Letters, that journal refused the second paper. It was subsequently 
accepted by Physical Review Letters after an anonymous referee
(generally held to be Nambu) suggested to Higgs that he emphasise
more the physical implications of his theory. Later in 1964, a more detailed
description of this idea appeared in a paper by Gerald Guralnik, Carl Hagen and Tom Kibble~\cite{GHK}.
Among all these 1964 papers, the only one to point out explicitly on
the appearance in the theory of a massive scalar boson was Higgs,
in his second paper~\cite{Higgs2}, which is why it is generally referred to as the Higgs
boson.

These authors considered the spontaneous breaking of an Abelian
U(1) gauge theory. The analogous phenomenon in a non-Abelian
theory was first studied by Sashas Migdal and Polyakov~\cite{MP}, who were unaware
of the earlier papers. Publication of their paper was delayed because
Soviet academicians could not believe that two young students, as
they were then, could come up with such a ground-breaking theory.
Partial breaking of a non-Abelian gauge symmetry was subsequently
rediscovered by Kibble~\cite{K}, and this is the form of spontaneous symmetry
breaking that is central to the Standard Model.

\subsection{Summary of the Standard Model}

\begin{table}
 \centering
 \begin{tabular}{|c|c|}
   \multicolumn{2}{c}{\textbf{Bosons}} \\
\hline
 \textbf{Gauge bosons} & \textbf{Higgs boson} \\ 
  \hline
  $\gamma$, $W^+$, $W^-$, $Z^0$, $g_{1 \ldots 8}$ & $\phi$ \\
  \hline 
\multicolumn{2}{c}{\textbf{Fermions}} \\
  \hline
  \textbf{Quarks} & \textbf{Leptons} \\
  \hline
  $\begin{array}{r}
   2/3: \\
   -1/3:
  \end{array} ~
  \left(\begin{array}{c}
   u \\ d       
  \end{array}\right) ~,
  \left(\begin{array}{c}
   c \\ s       
  \end{array}\right) ~,
  \left(\begin{array}{c}
   t \\ b       
  \end{array}\right)$ 
  &
  $\begin{array}{r}
   0: \\
   -1:
  \end{array} ~
  \left(\begin{array}{c}
   \nu_e \\ e^-       
  \end{array}\right) ~,
  \left(\begin{array}{c}
   \nu_\mu \\ \mu^-       
  \end{array}\right) ~,
  \left(\begin{array}{c}
   \nu_\tau \\ \tau^-       
  \end{array}\right)$ \\
  \hline
 \end{tabular}
 \vspace{2mm}
 \caption{\it Particle content of the Standard Model. Each quark comes in 3 colours, and the electric
 charges of the fermions are listed in the Table.}
\end{table}

Table~1 
summarizes the particle content of the Standard Model (SM)~\cite{S,W}.
The weak and electromagnetic interactions are described by a Lagrangian that is symmetric under gauge transformations
in a SU(2)$_L \times$U(1)$_Y$ group, where the subscript $L$ recalls that the weak SU(2) group 
acts only on left-handed fermions, and $Y$ is the hypercharge. We can write the 
SU(2)$_L \times$U(1)$_Y$ part of the SM Lagrangian as
\begin{eqnarray}\label{EqSMLag}
 \nonumber
 \mathcal{L} &=& -\frac{1}{4} {F}_{\mu\nu}^a {F}^{a\mu\nu} \\ \nonumber
             &+& i \overline{\psi} {/\mkern-13mu{D}}  \psi + h.c. \\ \nonumber
             &+& \psi_i y_{ij} \psi_j \phi + h.c. \\
             &+& \lvert D_\mu \phi \rvert^2 - V\left(\phi\right) ~,
\end{eqnarray}
which is short enough to write on a T-shirt or a pullover!

The first line in (\ref{EqSMLag}) contains the kinetic terms for the gauge bosons of the electroweak theory, 
where the index $a$ runs over the single U(1)$_Y$ gauge field, $\mathcal{A}_\mu$, and the
three gauge fields $W^{1,2,3}_\mu$ associated with SU(2)$_L$. The U(1) field-strength tensor
is the familiar 
\begin{eqnarray}
 f_{\mu\nu} &=& \partial_\nu \mathcal{A}_\mu - \partial_\mu \mathcal{A}_\nu ~,
 \label{EqFSTU1}
\end{eqnarray}
and the SU(2)$_L$ field-strength tensor is
\begin{eqnarray}
 F^a_{\mu\nu} &=& \partial_\nu W_\mu^a - \partial_\mu W_\nu^a + i g \epsilon_{abc} W_\mu^b W_\nu^c  \; \; \; {\rm for~~a = 1,2, 3} \, .
 \label{EqFSTSU2}
 \end{eqnarray}
where $g$ in (\ref{EqFSTSU2}) is the gauge coupling of SU(2)$_L$
and the $\epsilon_{abc}$ are its structure constants. The last term in (\ref{EqFSTSU2}) arises 
from the non-Abelian nature of the SU(2) group. 
At first sight, all the gauge fields are massless, in conflict with the massive nature of the weak bosons
$W^\pm$ and $Z^0$. As we see later, they acquire masses 
through the Englert-Brout-Higgs mechanism, whose physical manifestation is the Higgs boson.  

The second line in (\ref{EqSMLag}) contains the interactions between the spin-$1/2$ matter fields 
$\psi$ and the gauge fields via the covariant derivatives
\begin{equation}
 D_\mu = \partial_\mu + \frac{ig^\prime}{2} {A}_\mu Y + \frac{ig}{2} 
 {\tau} \cdot {W}_\mu ~,
 \label{covder}
\end{equation}
where $g^\prime$ is the U(1) coupling constant, $Y$ is the generator of the U(1) hypercharge, and
${\tau} \equiv \left(\tau_1, \tau_2, \tau_3\right)$ is the set of SU(2) Pauli matrices that represent the SU(2) algebra.

The third line in (\ref{EqSMLag}) describes the interactions between the matter fields 
and the Higgs field, $\phi$, via the Yukawa couplings $y_{ij}$, which give fermions their masses when 
the Higgs field acquires a vacuum expectation value (vev) $\langle \phi \rangle \ne 0$. 
In the SM the Higgs field $\phi$ is a complex doublet of SU(2) with non-zero U(1) hypercharge $Y$,
so this vev breaks electroweak symmetry.

The fourth and final line in (\ref{EqSMLag}) describes dynamics of the Higgs sector. The first term is the
kinetic term for the Higgs field, which also includes a covariant derivative $D_\mu$ (\ref{covder}),
and the second term in the final line of (\ref{EqSMLag}) is the Higgs potential $V(\phi)$:
\begin{equation}
V(\phi) \; = \; \ - \mu^2 |\phi|^2 + \lambda |\phi|^4 \, .
\label{Vphi}
\end{equation}
The negative sign of the first, quadratic term in (\ref{Vphi}) destabilizes the symmetric case $\langle \phi \rangle = 0$,
and the second, quartic term in (\ref{Vphi}) ensures that there is a stable minimum of the potential with 
\begin{equation}
\langle \phi \rangle \; \equiv \frac{v}{\sqrt{2}} \; = \; \frac{\mu}{\sqrt{2 \lambda}} \;  \ne \; 0 \, ,
\label{nonzerov}
\end{equation}
{\it if} $\lambda > 0$. The requirements that the coefficient of the quadratic term is negative and that of the quartic term is positive
are both problematic in the SM, as we shall see later.

Many different experiments have confirmed with high precision theoretical predictions 
derived from the first two lines in the SM Lagrangian (\ref{EqSMLag}).
However, until 2012 there was {no} experimental evidence for the last two lines, and there was considerable theoretical
doubt whether it could be correct. However, during Run~1 of the LHC the ATLAS~\cite{ATLASH} and CMS~\cite{CMSH}
Collaborations discovered a particle with properties resembling
those of the Higgs boson in the SM, as discussed later in this Lecture,
albeit with much less accuracy than, e.g., the precision electroweak tests
based on properties of the $W^\pm$ and $Z^0$ boson. The major tasks for future experiments at the LHC
and elsewhere will include probing whether the Higgs and other sectors of the SM Lagrangian in (\ref{EqSMLag})
hold up under more detailed scrutiny, whether there are additional interactions between SM particles, and
whether there is any evidence for new physics beyond the SM, as discussed in the second Lecture.

\subsection{Abelian (NG)AEBHGHKMP mechanism}

As a warm-up exercise, we consider the simplest Abelian model for spontaneous gauge symmetry breaking,
with just a U(1) gauge field $A_\mu$ and a single complex field $\phi$ described by the Lagrangian
\begin{equation}
{\cal L}(A, \phi) \; = \; -\frac{1}{4} {f}_{\mu\nu} {f}^{\mu\nu} +
(D_\mu \phi^\dagger )\left(D^\mu \phi \right) - V(\phi) \, ,
\label{Abelian}
\end{equation}
where ${f}_{\mu\nu}$ is given by ({\ref{EqFSTU1}), $D_\mu \equiv \partial_\mu - i e A_\mu$ 
and $V(\phi)$ has the `Mexican hat' form (\ref{Vphi})
illustrated in Fig.~\ref{fig:Mexhat}.
The U(1) gauge invariance implies that the theory is invariant under the local transformations
\begin{eqnarray}
\phi \to \phi^\prime & = & e^{i \alpha (x) } \phi \; = \;  e^{i \alpha (x) } e^{i \theta (x) } \eta (x) \, , \nonumber \\
A_\mu \to A_\mu^\prime & = & A_\mu (x) + \frac{1}{e} \partial_\mu \alpha (x) \, ,
\label{gauge}
\end{eqnarray}
where $\eta (x)$ and $\theta (x)$ are the magnitude and phase of $\phi(x)$, respectively.

\begin{figure}[ht]
\begin{center}
\includegraphics[width=10cm]{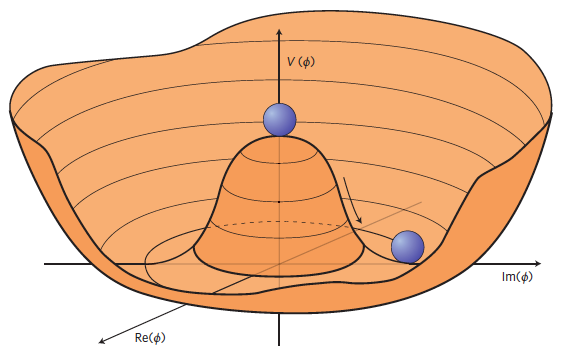}
\caption{\it The `Mexican hat' potential (\protect\ref{Vphi}).
The lowest-energy state may be described by a random point around the base of the hat. 
In the case of a global symmetry, motion around the bottom of the hat corresponds to a massless spin-zero
Nambu-Goldstone boson~\cite{Nambu,Goldstone}. 
In the case of a local (gauge) symmetry, 
this boson combines with a massless spin-one gauge boson to yield a
massive spin-one particle with three polarization states. The Higgs boson~\cite{Higgs2} is a 
massive spin-zero particle corresponding to quantum
fluctuations in the radial direction, up the side of the hat.}
\label{fig:Mexhat}
\end{center}
\end{figure}

We can exploit this gauge invariance to choose $\alpha (x) = - \theta (x)$, in which case $\phi^\prime (x) = \eta (x)$
and the Lagrangian (\ref{Abelian}) takes the form
\begin{equation}
{\cal L}(A_\mu^\prime, \eta) \; = \; -\frac{1}{4} {f^\prime}_{\mu\nu} {f^\prime}^{\mu\nu} + |(\partial_\mu - i e A_\mu^\prime) \eta |^2 - V(\eta) \, ,
\label{Abelianprime}
\end{equation}
The minimum of the potential $V (\eta)$ occurs at the following value of $\eta$:
\begin{equation}
\eta \; = \; \frac{v}{\sqrt{2}} \; \equiv \; \frac{\mu}{\sqrt{2 \lambda}} \, .
\label{vev}
\end{equation}
We can then rewrite (\ref{Abelianprime}) writing $\eta = (v + H)/\sqrt{2}$ and simplifying
the notation: $A_\mu^\prime \to A_\mu$, to obtain
\begin{equation}
{\cal L}(A_\mu, H) \; = \; -\frac{1}{4} {f}_{\mu\nu} {f}^{\mu\nu} + |(\partial_\mu - i e A_\mu)(\frac{v + H}{\sqrt{2}}) |^2 - V(\frac{v + H}{\sqrt{2}}) \, .
\label{Abelianeta}
\end{equation}
Expanding (\ref{Abelianeta}) to quadratic order in $A_\mu$ and $H$, we find
\begin{equation}
{\cal L}(A_\mu, H) \; \ni \; -\frac{1}{4} {f}_{\mu\nu} {f}^{\mu\nu} + e^2 v^2 A_\mu A^\mu + 
\frac{1}{2} [ (\partial_\mu H)^2 - m_H^2 H^2] + \dots \, , 
\label{Abelianquad}
\end{equation}
where the gauge boson has acquired a mass $m_A = ev/2$ and $m_H = \sqrt{2}\mu = \sqrt{2 \lambda} v$.
The mass of the vector boson results from the spontaneous symmetry gauge breaking mechanism,
with the phase degree of freedom $\theta$ of the complex scalar field $\phi$, which is a Nambu-Goldstone
boson, being `eaten' by the previously massless gauge boson - which has two polarization degrees of
freedom - to become the third polarization state needed for a
massive gauge boson.

The simultaneous appearance of a massive scalar boson $H$ is an inescapable feature of this mechanism,
since it is related to the positivity of the curvature in the scalar potential around the minimum,
which is needed to fix the vev $v$. This insight was made explicit in the second 1964 paper
by Higgs~\cite{Higgs2}, but was not mentioned in any of the other 1964 papers. 

\subsection{Spontaneous gauge symmetry breaking in the Standard Model}

The principle of spontaneous gauge symmetry breaking can easily be extended to the case of
a non-Abelian group, in particular the SU(2)$_L \times$U(1)$_Y$ of the Standard Model.
In this case, using the expressions (\ref{EqFSTU1}) and (\ref{EqFSTSU2}) for the gauge field
strengths and the expression (\ref{covder}) for the covariant derivative of what is now an isospin
doublet of Higgs fields $\phi$, we can expand the first (kinetic) term in the bottom line of
equation (\ref{EqSMLag}), $\lvert D_\mu \phi \rvert^2$, to obtain
\begin{equation}
{\cal L} \; \ni \; - \frac{g^2 v^2}{4} W_\mu^+ W^{\mu -} - \frac{g^{\prime 2} v^2}{8} B_\mu B^{\mu}
+ \frac{g g^\prime}{4} B_\mu W^{\mu 3} - \frac{g^2 v^2}{8} W_\mu^3 W^{\mu 3} + \dots \, .
\label{SMVmasses}
\end{equation}
where $v$ is the vev of the scalar field, which is determined in the same way as the previous
Abelian example.

The first term in (\ref{SMVmasses}) yields masses for the charged vector bosons $W^\pm$:
\begin{equation}
m_W \; = \; g \frac{v}{2} \, .
\label{mW}
\end{equation}
The other three terms yield a mass matrix for the two neutral gauge fields $(B, W^3)$,
which can be diagonalized to yield
\begin{eqnarray}
m_{Z} \; = \; \sqrt{g^2 + g^{\prime 2}} \frac{v}{2} \, & :  & Z \; \equiv \; \frac{g W^3 - g^\prime B}{\sqrt{g^2 + g^{\prime 2}}} \, , \nonumber \\
m_{A} \; = \; 0 \, & : & A \; \equiv \; \frac{g^\prime W^3 + g^\prime B}{\sqrt{g^2 + g^{\prime 2}}} \, .
\label{mZA}
\end{eqnarray}
The first of these mass eigenstates is the massive $Z$ studied in detail in experiments at the
LEP accelerator, in particular, and the second, massless eigenstate is identified with the photon.
It is useful to introduce the weak mixing angle $\theta_W$:
\begin{equation}
\tan \theta_W \; \equiv \; \frac{g^\prime}{g} \; \to \; m_Z \; = \; \frac{m_W}{\cos \theta_W}, \; Z \; = \; \cos \theta_W W^3 - \sin \theta_W B, \;
A \; = \; \sin \theta_W W^3 + \cos \theta_W A \, .
\label{thetaW}
\end{equation}
Measuring $\theta_W$ in different ways with high precision has provided important consistency
tests of the Standard Model, and provided a clue about the mass of the Higgs boson
before its discovery, as we discuss later.

As in the previous Abelian case, there is again a massive scalar (Higgs) boson whose mass is related to
the curvature of the potential $V(\phi)$ in the radial direction, and has the value
\begin{equation}
M_H \; = \; \sqrt{2} \mu \, .
\label{mH}
\end{equation}
The couplings of the Higgs boson to other Standard Model particles are predicted exactly.
Expanding the first (kinetic) term in the bottom line of
equation (\ref{EqSMLag}), $\lvert D_\mu \phi \rvert^2$, beyond quadratic order, we find
trilinear couplings of the Higgs boson to the massive gauge bosons:
\begin{equation}
g_{HWW} \; = \; \frac{2 m_W}{v}, \; \; g_{HZZ} \; = \; = \; \frac{2 m_Z^2}{v} \, ,
\label{VVH}
\end{equation}
and there are also important trilinear and quartic Higgs couplings.
As was first pointed out be Weinberg~\cite{W}, the third line in (\ref{EqSMLag}) links the Higgs-fermion couplings to their masses:
\begin{equation}
m_f \; = \; y_f v \; \leftrightarrow \; y_f \; = \; \frac{m_f}{v} \, .
\label{fH}
\end{equation}
The couplings (\ref{VVH}, \ref{fH}) lead to characteristic predictions for the partial
decay rates of the Standard Model Higgs boson:
\begin{equation}
\Gamma (H \to f {\bar f}) \; = \; N_c \frac{G_F m_H}{4 \pi \sqrt{2}} m_f^2 \, ,
\label{Gammaf}
\end{equation}
where the number of colours $N_c = 3$ for quarks, 1 for leptons,
and (for a sufficiently heavy Higgs boson)~\cite{Higgs3}
\begin{equation}
\Gamma (H \to W^+ W^-) \; = \; \frac{G_F m_H^3}{8 \pi \sqrt{2}} F( \frac{m_W}{m_H}) \, ,
\label{GammaV}
\end{equation}
where $F(m_W/m_H)$ is a phase-space factor, and there is a
corresponding formula for $\Gamma (H \to ZZ)$ with a prefactor $1/2$. Experimentally,
$m_H < 2 m_W$, so the Higgs boson cannot decay into pairs of on-shell gauge bosons,
but the decays $H \to W W^*, Z Z^*$ are quite distinctive, and have been measured.

Measurements of the couplings of the boson discovered in 2012 and checking their consistency
with the predictions (\ref{VVH}, \ref{fH}) have led to the general acceptance that it is indeed a
Higgs boson, as we also discuss later. Future higher-precision measurements will see
whether it is consistent with being the single Higgs boson of the Standard Model, or
whether the couplings exhibit deviations characteristic of some scenario for new physics 
beyond the Standard Model.

\subsection{Embarking on Higgs phenomenology}

In 1975 Mary Gaillard, Dimitri Nanopoulos and I made the first attempt at a systematic survey
of the possible phenomenological profile of the Higgs boson~\cite{EGN,Higgs3}. At that time, the Standard Model
was not established, idea of spontaneous gauge symmetry breaking was far from being 
generally accepted, there was general scepticism about scalar particles and, even if one 
bought all that, nobody had any idea how heavy a Higgs boson might be. For all these reasons,
we were rather cautious in the final paragraph of our paper, writing "we do not want to
encourage big experimental searches for the Higgs boson, but we do feel that people doing
experiments vulnerable to the Higgs boson should know how it may turn up."

Subsequently, searches for the Higgs boson were placed on the experimental agendas
of the LEP~\cite{Bjorn,LYR} and LHC accelerators at CERN. For example, a review of the possibilities
for new particle searches presented at the first workshop on prospective LHC physics
in 1984~\cite{EGK} discussed various ways of producing the Standard Model Higgs boson at the LHC.
There were also studies of Higgs production at the ill-fated SSC~\cite{EHLQ}, and the state of
play was described extensively in~\cite{HHG}.
However, in the 1980s there was still no indication what the Higgs mass might be.

The first clues about $m_H$ emerged from the high-precision measurements at LEP and the SLC
that started in 1989. These and other experiments found excellent overall agreement with the
predictions of the Standard Model. However, this consistency depended on the existences
if the top quark (which was discovered several years later) and the Higgs boson. The
accuracy of the LEP {\it et al.} measurements pointed towards (what seemed at that time) a
relatively heavy top quark~\cite{EF} and a Higgs boson weighing $\lesssim 180$~GeV~\cite{EFL}.

These indications came about through quantum (loop) corrections to electroweak observables,
such as the $W^\pm$ and $Z$ masses:
\begin{equation}
m_W^2 \sin^2 \theta_W \; = \; m_Z^2 \sin^2 \theta_W \cos^2 \theta_W \; 
= \; \frac{\pi \alpha}{\sqrt{2} G_F} \left( 1 + \Delta r \right) \; ,
\label{mWZ}
\end{equation}
where $\Delta r$ is the leading one-loop radiative correction, which exhibits the following
dependences on the top and Higgs masses:
\begin{equation}
\Delta r \; \ni \; \frac{3 G_F}{8 \pi^2\sqrt{2}} m_t^2 + \dots \; , \; \frac{G_F}{16 \pi^2} m_W^2 \left( \frac{11}{3} \ln \frac{m_H^2}{m_W^2} + \dots \right) \, ,
\label{Deltar}
\end{equation}
where we have exhibited the leading dependences on $m_t$ and $m_H$ for large masses.
These are relics of the divergences that would appear if either the top quark or the Higgs boson
were absent from the Standard Model, which would render it non-renormalizable.

In the early 1990s, even before the top quark was discovered, the precision electroweak data
were providing indications that the Higgs mass was probably well below the unitarity limit of $1$~TeV~\cite{unitarity},
which were strengthened when the top quark mass was measured~\cite{EFL}. Back in 2011,
just before the Higgs boson was discovered, the precision electroweak data suggested a
range $m_H = 100 \pm 30$~GeV. In parallel, unsuccessful searches at LEP had implied
that $m_H \ge 114$~GeV~\cite{LEPH}, and searches at the Fermilab Tevatron collider had excluded
a range around $(160, 170)$~GeV~\cite{TevatronH}. Combining all the information available in 2011, the
Gfitter Group obtained the $\chi^2$ likelihood function shown in Fig.~\ref{fig:Gfitter}~\cite{Gfitter},
corresponding to the estimate
\begin{equation}
m_H \; = \; 125 \pm 10 \; {\rm GeV} \, .
\label{estimated mH}
\end{equation}
The success of this prediction was a tremendous success for the Standard Model
at the quantum level.

\begin{figure}[ht]
\begin{center}
\includegraphics[width=10cm]{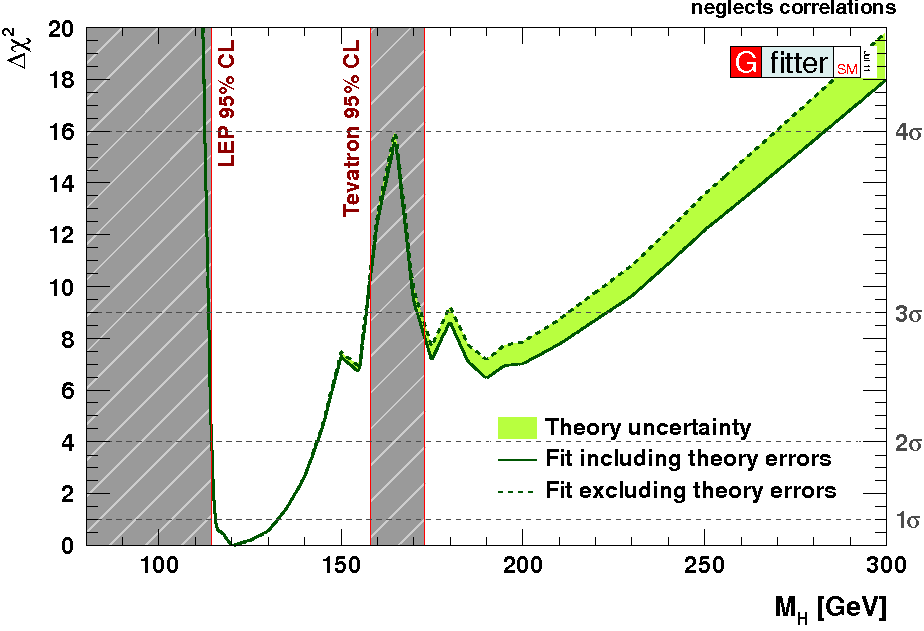}
\caption{\it The $\Delta \chi^2$ as a function of $m_H$ for a complete fit to the data
available in mid-2011~\protect\cite{Gfitter}, including precision electroweak data 
and the negative results of searches at LEP~\protect\cite{LEPH} and the
Fermilab Tevatron~\protect\cite{Gfitter} (grey bands). The solid (dashed) lines represent results that include (omit) theoretical
uncertainties.}
\label{fig:Gfitter}
\end{center}
\end{figure}

\subsection{Higgs production at the LHC}

Fig.~\ref{fig:LOH} displays various leading-order diagrams contributing to Higgs production at a proton-proton collider:
those for $gg \to H$, vector boson fusion and associated $V + H$ production are shown in the upper row, and
diagrams for associated ${\bar t}t + H$ and some of those for single $t + H$ production are shown in the lower row.
The left panel of Fig.~\ref{fig:HLHC} displays the most important Higgs production cross sections at the LHC
at 13~TeV in the centre of mass, as functions of the Higgs mass~\cite{LHCHXSWG4}. The dominant cross section
for $m_H \lesssim 1$~TeV is that for gluon fusion: $gg \to H$ via intermediate quark loops~\cite{GGMN}, 
the most important in the Standard Model being the top quark. The next most important
processes at low masses $m_H \lesssim 100$~GeV are the associated-production
mechanisms $q {\bar q} \to W + H, Z + H$~\cite{MNY}, whereas the vector-boson fusion processes
$W^+ W^-, ZZ \to H$~\cite{CD} are more important for $m_H \gtrsim 100$~GeV.
Next in the hierarchy of cross sections are the associated-production processes
$gg, q {\bar q} \to b {\bar b} H$ (which is difficult to distinguish) and  $t {\bar t} H$~\cite{RWNZK}
(which is more distinctive). Lowest in the hierarchy for $m_H \lesssim 200$~GeV is
the cross section for producing $H$ in association with a single $t$ or ${\bar t}$~\cite{Will}.
The right panel of Fig.~\ref{fig:HLHC} displays is a zoom of the cross sections
in a limited range of Higgs mass around 125~GeV~\cite{LHCHXSWG4}. The good news is that for 
$m_H \sim 125$~GeV most of these cross sections are potentially measurable 
at the LHC, and several of them have already been observed, as discussed later.

\begin{figure}[ht]
\begin{center}
\includegraphics[width=8cm]{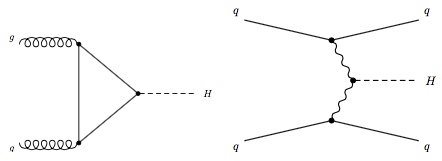}
\hspace{0.5cm}
\includegraphics[width=4cm]{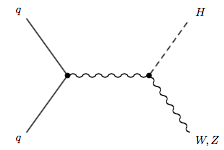} \\
\vspace{0.3cm}
~~~~~~~~~~~~~~~~~~~~~~~~~~~$gg \to H$~~~~~~~~~~~~~~~~~~Vector boson fusion~~~~~~ Associated $V+H$ production
\vspace{0.3cm}
\includegraphics[width=8cm]{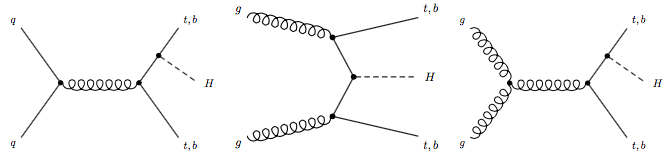}
\hspace{0.5cm}
\includegraphics[width=7cm]{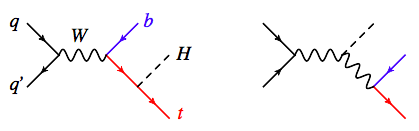} \\
~~~~~~~~~~~~~~~~~~~~~~~~${\bar t} t + H$ production~~~~~~~~~~~~~~~~~~~~~~~~~~~~~~$s$-channel diagrams for $t + H$ production
\vspace{0.5cm}
\caption{\it Leading-order diagrams for Higgs production. Upper row: $gg \to H$, vector boson fusion and associated $V+H$ production.
Lower row: ${\bar t} t + H$ production and $s$-channel diagrams for single $t + H$ production.}
\label{fig:LOH}
\end{center}
\end{figure}

\begin{figure}[ht]
\begin{center}
\includegraphics[width=8.5cm]{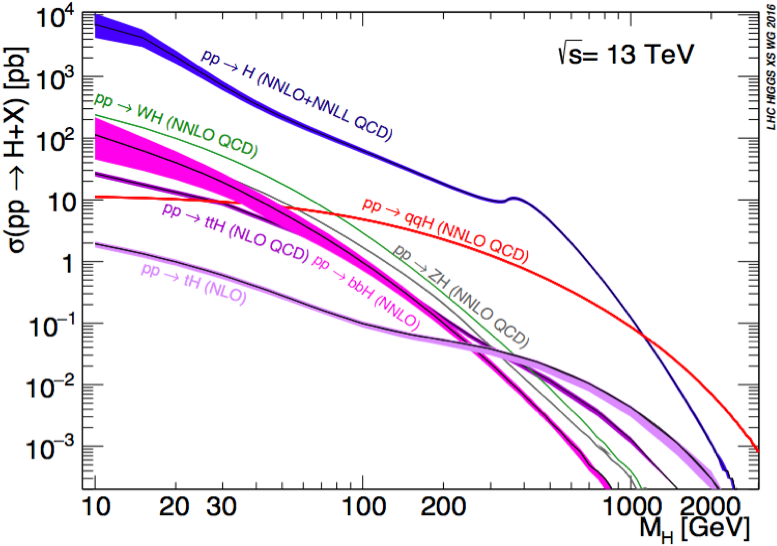}
\includegraphics[width=6.5cm]{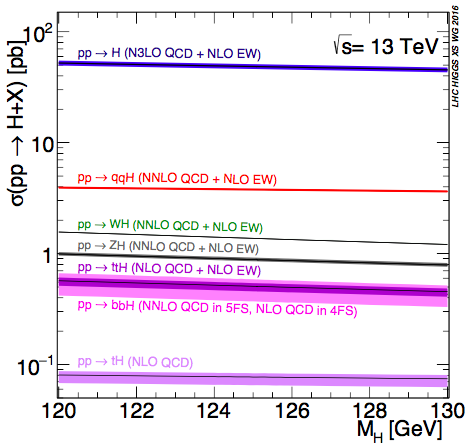}
\caption{\it Calculations of the dominant production cross sections for a
Standard Model Higgs boson, for a wide range of masses (left panel)
and for with a $m_H \in [120, 130]$~GeV (right panel)~\protect\cite{LHCHXSWG4}.
The uncertainties are represented by the widths of the coloured bands
for each production mechanism displayed.}
\label{fig:HLHC}
\end{center}
\end{figure}

As can be seen in Fig.~\ref{fig:HLHC}, the dominant $gg \to H$ cross section has a relatively 
large uncertainty. This is because it is a strong-interaction process, which has relatively
large perturbative corrections, and is induced at the loop level, implying that the calculation of
these corrections is more arduous than for a tree-level process. Nevertheless, a complete
calculation of the $gg \to H$ cross section at the next-to-next-to-leading order (NNLO)
is available, as is a heroic N$^3$LO calculation in the limit of a heavy top quark~\cite{heroic}. The
result of these efforts is the following estimate of the $gg \to H$ cross section~\cite{LHCHXSWG4}:
\begin{equation}
\sigma \; = \; 48.58 \; ^{+2.22}_{-3.27} \; {\rm (theory)} \; \pm 1.56 \; {\rm (PDF,} \; \alpha_s) \; {\rm pb} \, ,
\label{sigmaggH}
\end{equation}
corresponding to a total uncertainty of $< 10$\%. It is worth noting that the NLO
correction more than doubled the cross section, that the the NNLO correction was
about 20\% of the final estimate (\ref{sigmaggH}). However, the N$^3$LO correction
was only $\sim 3$\%, indicating that the perturbative QCD corrections are under control.
Table~2
compiles the principal theoretical uncertainties in the calculation
(\ref{sigmaggH}) of the $gg \to H$ cross section. The first is associated with the choice of
scale in the perturbative QCD calculation, and the second is an estimate of the
uncertainty due to the truncation of the perturbative expansion. The third is an estimate
of the theoretical uncertainty in the use of the parton distribution functions (PDFs) and $\alpha_s$, and the
fourth is an estimate of the uncertainty in higher-order mixed electroweak and QCD
perturbative corrections. The fifth is the parametric uncertainty in the values of
$m_{t, b, c}$ to be used, and the sixth and last is an estimate of the uncertainty
in the heavy-top approximation in the N$^3$LO calculation. We see that many
uncertainties are comparable at the level of $\pm {\cal O}(0.5)$~pb, indicating that
a struggle on many fronts will be needed to reduce substantially the theoretical error
in (\ref{sigmaggH}).

\begin{table}
 \centering
 \begin{tabular}{|c|c|c|c|c|c|}
 \hline
 Scale & Truncation & PDFs (TH) & Electroweak & $m_{t, b, c}$ & $1/m_t$ \\
 \hline
 $^{+0.10}_{-1.15}$ & $\pm 0.18$ & $\pm 0.56$ & $\pm 0.49$ & $\pm 0.40$ & $\pm 0.49$ \\
  \hline
 \end{tabular}
 \vspace{2mm}
 \caption{\it Breakdown of the theoretical uncertainties (in pb) in (\protect\ref{sigmaggH})
 that are associated with different
 approximations in the calculation of the $gg \to H$ cross section~\protect\cite{LHCHXSWG4}.}
\end{table}

Concerning the PDF uncertainties in (\ref{sigmaggH}), Fig.~\ref{fig:PDFs} shows that
there is now good consistency between the $gg$ collision luminosities estimated
by different PDF fitting groups with the recommendation of the PDF4LHC Working
Group~\cite{PDF4LHC}. The uncertainty from this source is currently estimated at $\sim 2$\%,
which is comparable to the parametric uncertainty associated with $\alpha_s$. These
are currently the largest individual sources of uncertainty in the $gg \to H$ cross section.

\begin{figure}[ht]
\begin{center}
\includegraphics[width=12cm]{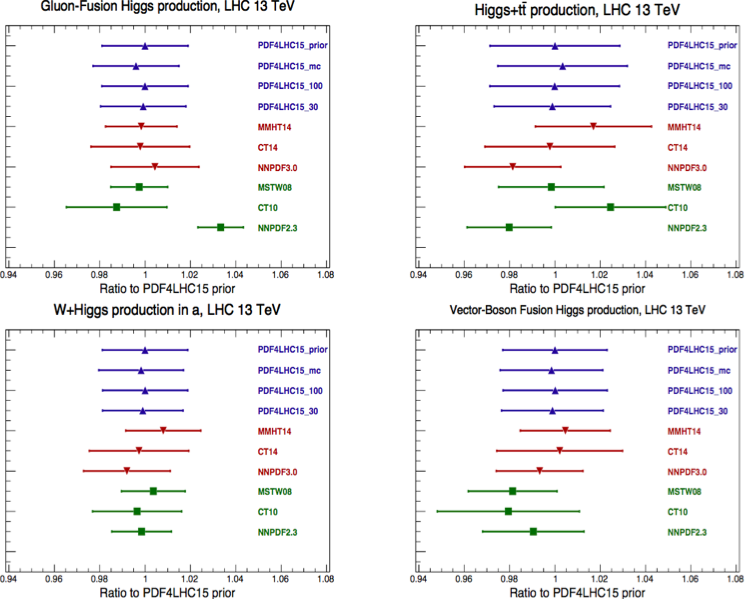}
\caption{\it Comparison between the parton-parton production luminosities calculated
using different PDF sets~\protect\cite{LHCHXSWG4}, compared to the PDF4LHC recommendation~\protect\cite{PDF4LHC}.}
\label{fig:PDFs}
\end{center}
\end{figure}

There are smaller uncertainties in the cross sections for vector boson fusion (shown in the upper panels of Fig.~\ref{fig:otherH})
and $H$ production in association with a $W^\pm$ or $Z$ boson (lower left panel of Fig.~\ref{fig:otherH}), both of which
have been calculated at NNLO including electroweak corrections at NLO~\cite{LHCHXSWG4}. In both cases,
there is good convergence of the perturbation expansion, and there are also quite
small uncertainties in the relevant quark parton PDFs.
On the other hand, the uncertainties in the cross section for associated $t {\bar t} + H$
production (shown in the lower right panel of Fig.~\ref{fig:otherH})
are significantly greater. This is a strong interaction cross section that has
been calculated only at NLO, so there are considerable uncertainties associated with
the perturbation expansion. Also, there are greater uncertainties associated with the
choices of quark, antiquark and gluon PDFs. The situation is similar for single
$t$ or ${\bar t} + H$ associated production.

\begin{figure}[ht]
\begin{center}
\includegraphics[width=12cm]{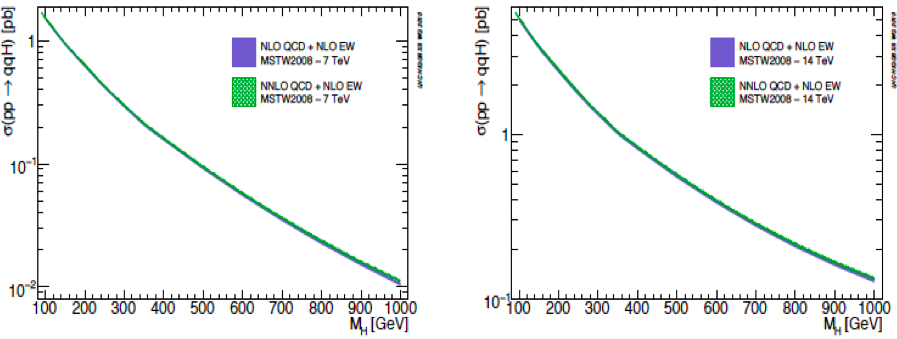} \\
\vspace{0.5cm}
\hspace{-0.35cm}
\includegraphics[width=12.5cm]{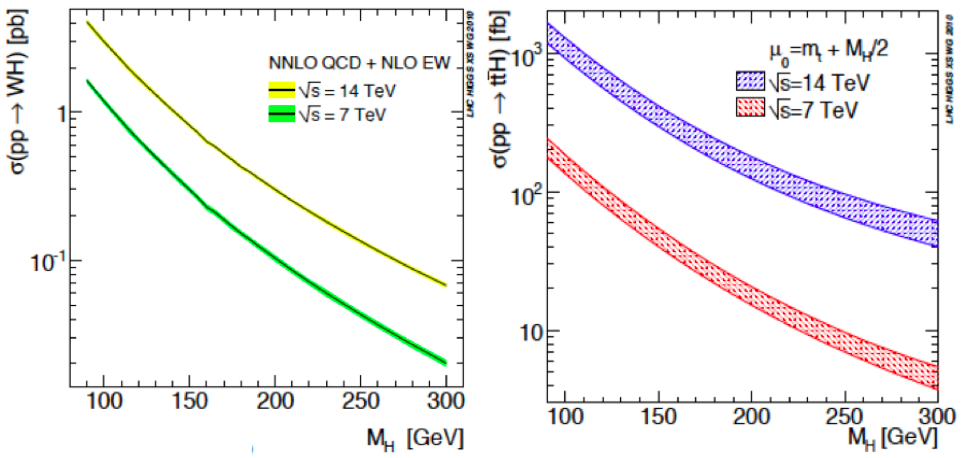}
\caption{\it The cross sections for vector boson fusion production of the Higgs boson with the LHC at 7 TeV (upper left panel),
and at 14 TeV (upper right panel), for associated $WH$ production at 7 and 14 TeV (lower left panel), and for
${\bar t}t H$ production at 7 and 14 TeV (lower right panel), all as functions of $m_H$~\protect\cite{LHCHXSWG3}.}
\label{fig:otherH}
\end{center}
\end{figure}

\subsection{Higgs decays}

Since Higgs couplings to other particles are proportional to their masses,
as seen in (\ref{VVH}) and (\ref{fH}), it is expected to decay predominantly
into the heaviest particles that are kinematically accessible~\cite{EGN}. 
This is apparent
in the left panel of Fig.~\ref{fig:Hdecays}, which gives an overview of Higgs decay
branching ratios for a wide range of Higgs masses. In the specific case of interest when 
$m_H \simeq 125$~GeV, as seen in the right panel of Fig.~\ref{fig:Hdecays}~\cite{LHCHXSWG4},
we see that $H \to b {\bar b}$ decays are expected to dominate, with
$H \to c {\bar c}$ decays suppressed by $(m_c/m_b)^2$, and $H \to \tau^+ \tau^-$
decays suppressed by a missing colour factor as well. Since $m_H < 2 m_{W, Z}$,
only the off-shell decays $H \to W W^*, Z Z^*$ followed by $W^*, Z^* \to f {\bar f}$
decays are possible. Nevertheless, because of the large vector-boson mass factors in (\ref{VVH}),
these three-body decays have branching ratios comparable to the leading 
$H \to f {\bar f}$ two-body decays, as seen in the right panel of Fig.~\ref{fig:Hdecays}.

\begin{figure}[ht]
\begin{center}
\includegraphics[width=8.9cm]{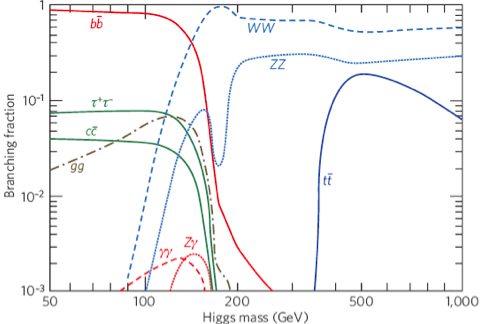}
\includegraphics[width=6.4cm]{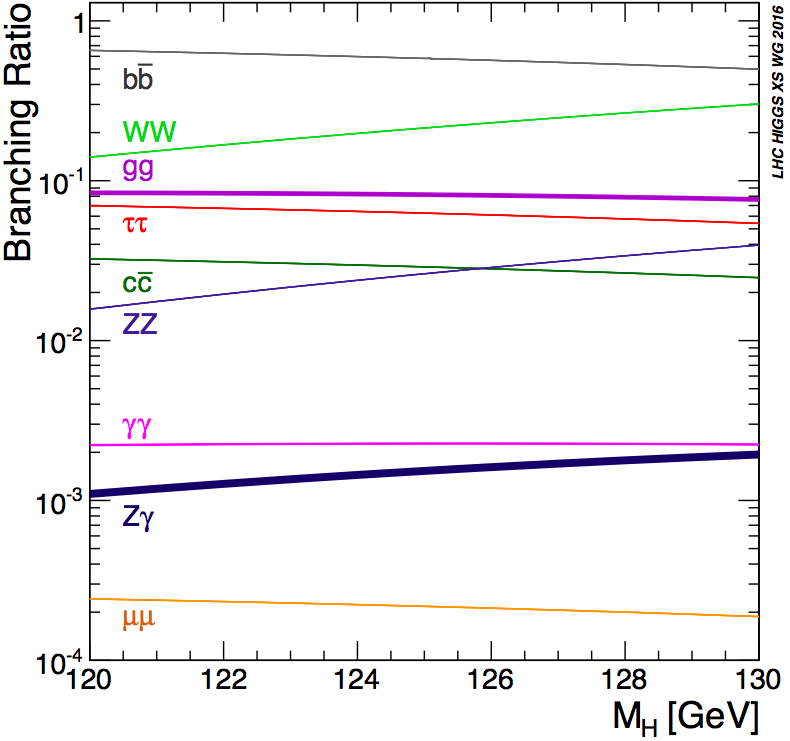}
\caption{\it Calculations of the dominant decay branching ratios for a
Standard Model Higgs boson over a large range of masses (left panel)
and with mass $m_H \in [120, 130]$~GeV (right panel)~\protect\cite{LHCHXSWG4}, where
the uncertainties are represented by the widths of the coloured bands
for each decay mode displayed.}
\label{fig:Hdecays}
\end{center}
\end{figure}

Also comparable is the rate for $H \to gg$ decay, which is a loop-induced process:
see Fig.~\ref{fig:LOgg}.
It is difficult to observe $H \to gg$ decay directly, but is related to the rate for the dominant
$gg \to H$ fusion production process. The electroweak loop-induced decays
$H \to \gamma \gamma$ and $Z \gamma$ occur with somewhat lower branching
ratios. However, the $H \to \gamma \gamma$ mode is very clean, and was one of
the discovery modes at the LHC. 
The amplitude for $H \to \gamma \gamma$ decay is generated by loops of
massive charged particles~\cite{EGN} as shown in Fig.~\ref{fig:LOgg}.
The most important contributors in the Standard Model are the top quark and
the $W^\pm$ boson. Their contributions may be written as follows:
\begin{equation}
\Gamma (H \to \gamma \gamma) \; = \; \frac{G_F \alpha^2 m_H^3}{128 \pi^3 \sqrt{2}}
\left| \Sigma_f N_c Q_f^2 A_{1/2} (r_f) + A_1 (r_W) \right|^2 \, ,
\label{bgammadecay}
\end{equation}
where $A_{1/2}$ and $A_1$ are known functions of $r_f \equiv m_f / m_H$ and 
$r_f \equiv m_W / m_H$ that have opposite signs, so that
the top and $W^\pm$ contributions interfere destructively. In the Standard Model,
the $gg H$ amplitude receives contributions only from the top and (less important)
lighter quarks.

The last decay mode shown in the right panel of
Fig.~\ref{fig:Hdecays} is $H \to \mu^+ \mu^-$, which is suppressed relative to
$H \to \tau^+ \tau^-$ by a factor $(m_\mu/m_\tau)^2$. Nevertheless, it also has
quite a clean experimental signature, and is expected to be the first decay of $H$
into second-generation fermions to be probed.

\begin{figure}[ht]
\begin{center}
\includegraphics[width=12cm]{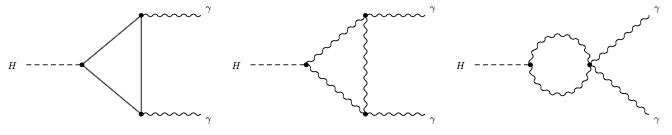}
\caption{\it Loop diagrams for $H$ couplings to massless gauge bosons. In the Standard Model,
the dominant fermion diagram (left) for $H \to gg$ and $\gamma \gamma$ involves the top quark.
The $W^\pm$ diagrams (middle and right) contribute only to $H \to \gamma \gamma$. There are similar
diagrams for $H \to Z \gamma$.}
\label{fig:LOgg}
\end{center}
\end{figure}

There have been strong theoretical efforts to calculate perturbative corrections to
$H$ decays~\cite{LHCHXSWG4}, leading to the relatively small uncertainties shown in
Fig.~\ref{fig:Hdecays}. The largest relative uncertainties are in $H \to gg$ decay,
because it is a loop-induced decay into strongly-interacting particles, and 
$H \to Z \gamma$ decay, where high-order calculations are complicated by the
masses of the initial-state $H$ and the final-state $Z$.

Echoing what was said earlier about Higgs production mechanisms, another piece
of good news is that for $m_H \simeq 125$~GeV many Higgs decay modes ares
measurable at the LHC. These happy chances provide many opportunities to
measure distinctive properties of the Higgs boson. Also, it is an amusing irony that 
the largest $H$ production mechanism, $gg \to H$, and one of the cleanest $H$ 
decay channels, $H \to \gamma \gamma$, are both loop-induced processes. Thus, 
LHC data already give us access to quantum aspects of Higgs physics, including
the possible existence of new heavy particles beyond the Standard Model.

If the Higgs boson had weighed 750~GeV (just saying)~\cite{750}, gathering a lot of
information about it would have been more difficult. In that case, observing other
production modes besides $gg \to H$ and vector-boson fusion at the LHC would 
have been difficult, and observing any other decays besides $H \to W^+ W^-, ZZ$ 
and $t {\bar t}$ would probably have been impossible in the absence of any other 
new particles~\cite{EEQSY}. Obviously, we regret the passing of the late lamented $X(750)$
particle, which would have required new physics to explain its production and
decay, but we should thank our lucky stars for the openness of the $H(125)$!

\subsection{The stakes in the Higgs search}

The stakes in the Higgs search were very high. How is gauge symmetry broken:
spontaneously (elegantly) or explicitly (ugly and uncalculably)? Assuming that
it is broken spontaneously, is it broken by an elementary scalar field, which would
be a novelty that raises perhaps more questions than it answers, many of which
are related to the hierarchy of mass scales in physics? The Higgs is very likely a
portal towards many issues in physics beyond the Standard Model. It would have been
associated with a phase transition in the Universe when it was about $10^{-12}$
seconds old, which maight have been when the baryon asymmetry of the
Universe was generated. The Higgs or a related scalar field might have caused
the Universe to expand (near-)exponentially in a bout of cosmological inflation
when it was about $10^{-35}$ seconds old. And a Higgs field should contribute
a factor $\sim 10^{60}$ too much to the dark energy measured in the Universe
today. The stakes in the search for the Higgs boson were undoubtedly high!

\subsection{The mass of the Higgs boson}

The discovery of the Higgs boson in 2012~\cite{ATLASH,CMSH} was primarily based on the observation
of excesses of events in the $\gamma \gamma$ and $2 \ell^+ 2 \ell^-$ channels
(where $\ell = \mu$ or $e$), interpreted as being due to $H \to Z Z^*$, together with a 
broad excess of $\ell^+ \ell^-$ + missing transverse energy events, interpreted as
being due to $H \to W W^*$.

Measurements of the $\gamma \gamma$ and $2 \ell^+ 2 \ell^-$ final states have
enabled the mass of the Higgs boson to be determined with high precision. The final
combined results from ATLAS and CMS LHC Run 1 data yield~\cite{ATLAS+CMS}
\begin{equation}
m_H \; = \; 125.09 \pm 0.21 \; ({\rm statistical}) \pm 0.11 \; ({\rm systematic}) \; {\rm GeV} \, ,
\label{measuredmH}
\end{equation}
a measurement at the level of 2 {\it per mille} that is dominated by the statistical error
and hence can be further reduced. An accurate measurement of $m_H$ is a {\it sine
qua non} for precision tests of the Standard Model, since it enters in the Higgs production
cross section and decay branching ratios, as seen in the right panels of 
Figs.~\ref{fig:HLHC} and \ref{fig:Hdecays}. Moreover, it is crucial for the discussion
below of the stability of electroweak vacuum. As already mentioned, the measurement
(\ref{measuredmH}) is fully in line with previous indications from precision electroweak
data and previous searches at LEP~\cite{LEPH} and the Fermilab Tevatron collider~\cite{TevatronH}, see
Fig.~\ref{fig:Gfitter}. Fig.~\ref{fig:GfitterEW} shows a direct comparison between the
ATLAS and CMS measurements of $m_H$ and the $\chi^2$ function from a global
analysis of the precision electroweak data, omitting the LEP and Tevatron constraints~\cite{GfitterEW}.
We see that the measured value of $m_H$ agrees with the indication from the
electroweak data at the $\Delta \chi^2 \sim 1.5$ level.

{\it This may seem like a disaster for the quest for physics beyond the Standard Model,
but not so fast!}

\begin{figure}[ht]
\begin{center}
\includegraphics[width=10cm]{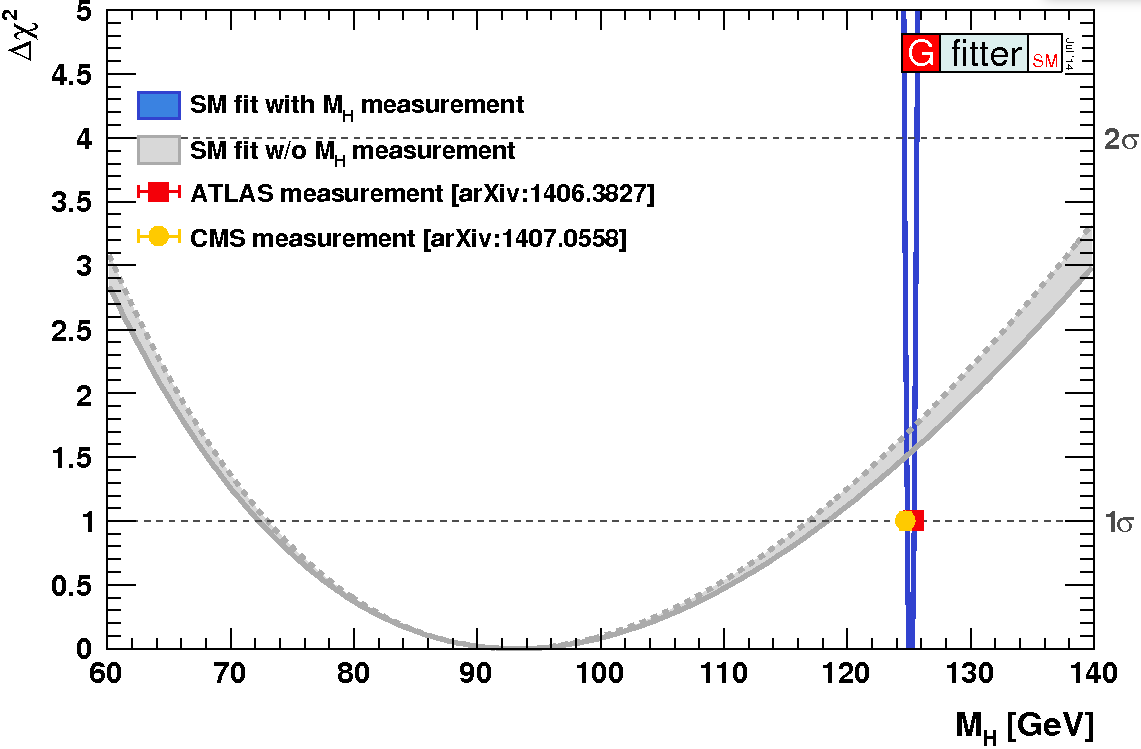}
\caption{\it The $\Delta \chi^2$ as a function of $m_H$ for a fit to the precision electroweak data
compared with the measurements of $m_H$ by the ATLAS and CMS Collaborations~\protect\cite{GfitterEW}.}
\label{fig:GfitterEW}
\end{center}
\end{figure}

\subsection{The instability (?) of the electroweak vacuum}

Let us consider the second, $\lambda \phi^4$, term in the Higgs potential (\ref{Vphi})
It is essential for the Mexican hat form of the potential seen in Fig.~\ref{fig:Mexhat}
and the existence of the non-zero vev (\ref{nonzerov}) that $\lambda > 0$. However,
like any other coupling in a quantum field theory, $\lambda$ is subject to
renormalization. In the Standard Model, there are two important sources of this renormalization at the
one-loop level. One is that due to $\lambda$ itself, which tends to {\it increase}
$\lambda$ as the renomalization scale $Q$ increases:
\begin{equation}
\lambda (Q) \; \simeq \; \frac{ \lambda (v)}{1 - \frac{3}{4 \pi^2} \lambda (v) \ln \left(Q^2/v^2 \right)} \, .
\label{lambdalambda}
\end{equation}
Left to itself, this self-renormalization would cause $\lambda$ to blow up at some high
renormalization scale $Q$. However, there is also importantant one-loop renormalization of 
$\lambda$ due to loops of top quarks:
\begin{equation}
\lambda (Q) \; \simeq \; \lambda (v) - \frac{3 m_t^4}{4 \pi^2 v^2} \ln \left(Q^2/v^2 \right) \, ,
\label{lambdatop}
\end{equation}
which tends to {\it decrease} $\lambda$ as the renomalization scale $Q$ increases,
driving it towards negative values. If $\lambda$ indeed turns negative, there soon appears a
field value with lower energy than our electroweak vacuum, which becomes unstable
or at least metastable.

The left panel of Fig.~\ref{fig:unstable} illustrates how the negative renormalization by the
top quark drives $\lambda < 0$ in the Standard Model~\cite{DDEEGIS,BDGGSSS}, though this is subject to uncertainties
in $m_t$, in particular. As seen in the right panel of Fig.~\ref{fig:unstable}, the current world
averages of $m_t$ and $m_H$ suggest that these parameters indeed lie within the region
where the Standard Model electroweak vacuum is metastable.
{\it In my view, this is a potential disaster (pun intended) that would require new physics
to avert it.}

\begin{figure}[h!]
\centering
\includegraphics[width=7.1cm]{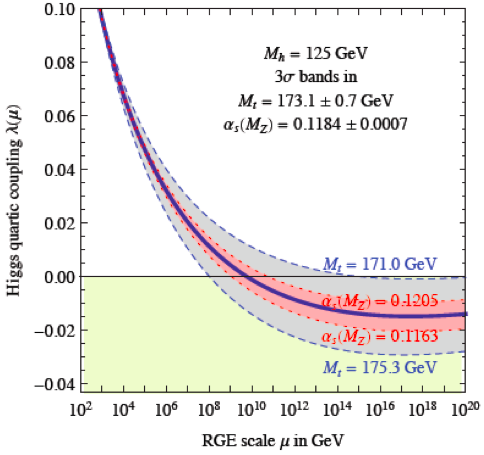}
\includegraphics[width=7cm]{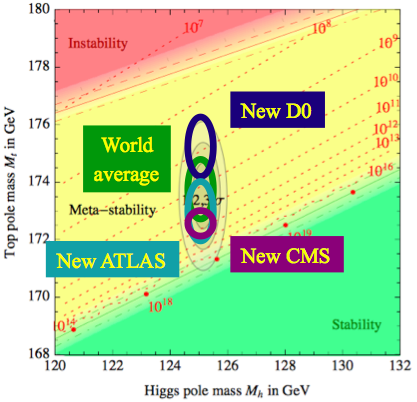}
\caption{\it Left panel: The negative renormalization of the Higgs self-coupling by the top quark within the Standard Model
leads to an instability in the Higgs potential for field values $\sim 10^{9}$~GeV\protect\cite{DDEEGIS}.
Right panel: Experimental measurements of $m_t$ and $m_H$ suggest that the electroweak vacuum of the Standard
Model would be metastable, modulo uncertainties in $m_t$, in particular~\protect\cite{BDGGSSS}.}
\label{fig:unstable}
\end{figure}

As seen in Fig.~\ref{fig:unstable}, the location and indeed existence of the instability scale $\Lambda_I$ are
particularly sensitive to $m_t$, and also to $\alpha_s$ as well as to $m_H$. One calculation including higher-order effects yields
the following dependences on these parameters~\cite{BDGGSSS}:
\begin{equation}
\log_{10} \left( \frac{\Lambda_I}{\rm GeV}\right) \; = \; 9.4 + 0.7 \left(\frac{m_H}{\rm GeV} -125.15 \right)
- 1.0 \left( \frac{m_t}{\rm GeV} - 173.34 \right) + 0.3 \left( \frac{\alpha_s(m_Z) - 0.1184}{0.0007} \right) \, .
\label{LambdaI}
\end{equation}
Inserting the world average value (\ref{measuredmH}) for $m_H$, $m_t = 173.3 \pm 1.0$~GeV
and $\alpha_s (m_Z) = 0.1181 \pm 0.0011$, we estimate
\begin{equation}
\log_{10} \Lambda_I) \; = 9.4 \pm 1.1 \, ,
\label{LambdaIvalue}
\end{equation}
indicating that (in the immortal words of the Apollo 13 astronauts) we have a problem~\cite{others}.

Several words of caution are in order. The first is that the experimental measurement
of $m_H$ (\ref{measuredmH}) is already so accurate that it is the smallest source of uncertainty
in (\ref{LambdaI}). Concerning the larger uncertainty due to $m_t$, subsequent to the compilation 
of the world average value~\cite{WAmt}, several new measurements have been published, including by D0:
$m_t = 174.98 \pm 0.76$~GeV~\cite{D0mt}, by ATLAS: $m_t = 172.99 \pm 0.91$~GeV~\cite{ATLASmt}
and by CMS: $m_t = 172.4 \pm 0.49$~GeV~\cite{CMSmt}. These scatter above and below
the official world average, and a new compilation awaits better understanding of the
discrepancies between them. However, all these measurements lie in the unstable
region of $m_t$. A second comment concerns the interpretation of the value of $m_t$
that the experiments measure. This is defined within a specific Monte Carlo event simulation
code, and one issue concerns the relation between this and the pole mass. Present
understanding is that the difference between these definitions $\sim \Lambda_{QCD} < 1$~GeV,
so this correction seems manageably small at the present time, though it will require more
detailed analysis as the experimental precision improves. Another issue concerns the
relation between the pole mass and the $\overline{MS}$ mass that is used in the loop
calculations of vacuum stability. This relation has been calculated to ${\cal O}(\alpha_s^4)$~\cite{mt4}:
\begin{eqnarray}
\frac{m_t|_{\rm pole} - m_t|_{\overline{\rm MS}}}{m_t|_{\overline{\rm MS}}} \; \equiv \; \frac{\delta m}{m_t|_{\overline{\rm MS}}}
& \equiv & \Delta \; = \; 0.4244 \, \alpha_s (m_t)
\, + \, 0.8345 \, \alpha_s (m_t)  \nonumber \\
& & \quad \, + \, 2.375 \, \alpha_s (m_t)^3 \, + \, (8.49 \pm 0.25) \, \alpha_s (m_t)^4 + \dots \, ,
\label{QCD4}
\end{eqnarray}
leading to the numerical result
\begin{equation}
m_t|_{\rm pole} \; = \; m_t|_{\rm \overline{MS}} + 7.557 + 1.1617 + 0.501 + 0.195 + \dots \, ,
\label{mtrelation}
\end{equation}
where the numbers are for the ${\cal O}(\alpha_s^n)$ corrections calculated
for $n = 1,2, 3, 4$ assuming $m_t|_{\overline{MS}} = 163.643$~GeV. These corrections
decrease systematically in magnitude, so the QCD perturbation series seems to be
well-behaved. The sum of the uncalculated higher-order terms has been estimated
to be $\sim 250$~GeV, with uncertainty of $\sim 70$~GeV~\cite{Nason}. These effects
are also below within the current experimental precision.

Another comment concerns the length of the lifetime of our (in principle) unstable electroweak vacuum,
which may be much longer than the age of the Universe to date. This is certainly a
necessary consistency condition for the existence of physicists capable of
recognizing the problem, but also leads some of them to disregard it as unimportant.
I disagree with this attitude for two reasons. One is because the present vacuum energy
is very small and positive. Arguments have been proposed how this might come about
if our vacuum is (one of) the lowest-energy state(s) in an extensive landscape,
or one might imagine that some approximate symmetry could yield a small positive value.
Personally, I would find the small value of the present vacuum energy much more difficult to understand
if it is only a temporary state, and
if our universe will eventually decay into an anti-De Sitter state with negative vacuum
energy of much larger magnitude. The other reason for taking vacuum stability seriously
as a requirement is that, if it were not, fluctuations in the Higgs field in the hot and dense
early universe would have taken most of it into the anti-De Sitter ``Big Crunch" phase, and
the conventional expansion of the universe would never have occurred~\cite{FGH,Hook}. On the other hand, one
could argue anthr*p*c*lly that if even an infinitesimal part of the universe escaped the
``Big Crunch", that would have been enough for sentient physicists to come into being.

My own take on the instability problem is that we should take it seriously, and that it
motivates some form of new physics to stabilize the electroweak vacuum~\footnote{Assuming
that we have not misunderstood the experimental measurements of $m_t$.}. Clearly, any
such physics should appear at some energy scale below $\Lambda_I \sim 10^{9}$~GeV,
but might lie far beyond the reach of conceivable accelerators. However, in my view it is the best
hint for some new physics beyond the Standard Model provided by Run 1 of the LHC.

What might this new physics look like? As we saw above, it is the negative sign of the top quark 
loop that destabilizes the effective Higgs potential. In order to counteract it, one
should introduce a scalar $\phi$, whose loop would have a positive sign, and
which could in general have couplings of the forms~\cite{ER}:
\begin{equation}
{\cal L} \; \ni \; M^2 |\phi|^2 + \frac{M_0^2}{v^2} |H|^2 |\phi|^2 \, ,
\label{ER1}
\end{equation}
where $M$ and $M_0$ are two mass parameters. Indeed, if one chooses
$M \lesssim 10^5$~GeV, the effective Higgs potential can be stabilized. However,
avoiding a blow-up in $\lambda$ as well as a negative value typically requires
some fine-tuning of $M_0$, i.e., the coupling of the new scalar $\phi$ to the
Higgs field $H$, at the {\it per mille} level. The simplest way to stabilize the
coupling is to postulate new fermions to counteract the $\phi - H$ coupling in (\ref{ER1}).
But now we have introduced scalar partners of the top and fermionic partners
of the Higgs that make the theory reminiscent of supersymmetry~\cite{ER}.
So, why not postulate supersymmetry, as we discuss in the second Lecture?

\subsection{Higgs coupling measurements}

The ATLAS and CMS Collaborations have published a joint analysis of their
measurements of Higgs production and decay in various channels~\cite{ATLAS+CMScouplings}, as shown in 
Fig.~\ref{fig:Hcouplings}. Several Higgs decay modes have been established
with high significance, including $\gamma \gamma$, $W W^*$, $Z Z^*$ and $\tau^+ \tau^-$,
and there are important constraints on other Higgs decay models.
The $gg \to H$ and vector boson fusion production mechanisms have been
established, and there are interesting constraints on $H$ production in association
with $W^\pm$, $Z$ and $t {\bar t}$ pairs.

\begin{figure}[ht]
\begin{center}
\includegraphics[width=7cm]{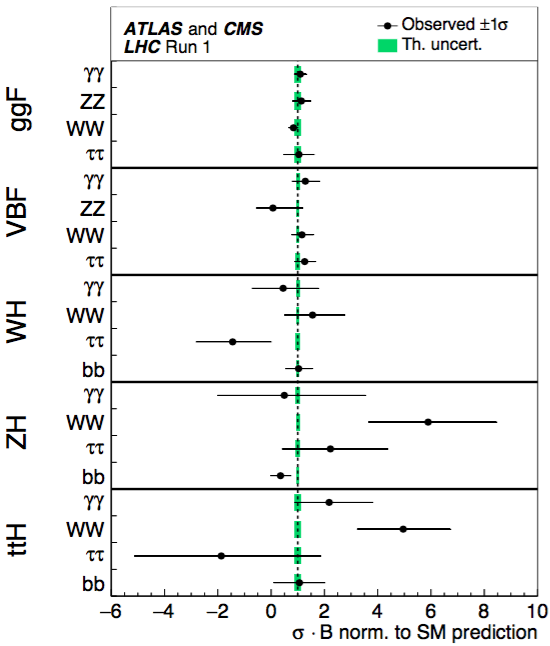}
\caption{\it The products of cross sections $\sigma$ and branching ratios B
measured by the ATLAS and CMS Collaborations in various channels,
normalized to the Standard Model predictions~\protect\cite{ATLAS+CMScouplings}.}
\label{fig:Hcouplings}
\end{center}
\end{figure}

However, many $H$ couplings remain to be established. Most prominently,
the $H \to b {\bar b}$ decay that is expected to dominate has been seen only
at the 2.6-$\sigma$ level at the LHC and the 2.8-$\sigma$ level at the Tevatron~\cite{CDF+D0H}.
Moreover, although there is indirect evidence for a $t {\bar t} H$ coupling
from measurements of the induced $ggH$ and $\gamma \gamma H$ couplings,
there is no direct evidence in the absence of measurements of $t {\bar t} H$
(or single $t/{\bar t} H$) associated production. Also, there is as yet no evidence
for $H \to \mu^+ \mu^-$ decay, though upper limits on it  already provide
interesting information, as we see in the second Lecture.

The first analyses of Higgs data from Run 2 of the LHC have also been shown
by ATLAS and CMS~\cite{Run2H}. The distinctive $H \to \gamma \gamma$ and $Z Z^*$ decays
have been seen again with 10-$\sigma$ significance, the Higgs production
cross section at 13 TeV is in line with theoretical calculations, and the searches are on
for $H \to \mu^+ \mu^-$ and associated $t {\bar t} H$ and single $t/{\bar t} H$ production.

\subsection{Not a big deal?}

Shortly after the discovery of the Higgs boson, Peter Higgs was quoted in the Times
of London as saying: ``A discovery widely acclaimed as the most important scientific
advance in a generation has been overhyped"~\cite{HiggsTimes}. I would very humbly and respectfully beg to disagree.
Without the Higgs boson (or something to do its job), there would be no atoms
because electrons would escape from nuclei at the speed of light, the weak interactions
responsible for radioactivity would not be weak, and the universe would be totally
unliveable. {\it It was a big deal.}

\section{The Particle Physics Higgsaw Puzzle}

\subsection{Has the LHC found the missing piece?}

The first piece of the particle physics jigsaw puzzle to be discovered was the electron
in 1997, and it took 115 years until a candidate for the final missing piece of the
Standard Model, the Higgs boson, was discovered in 2012~\cite{ATLASH,CMSH}. In the first Lecture,
I was jumping the gun, blithely assuming that the particle discovered in 2012 is 
indeed the Higgs boson. In this Lecture we first review the experimental and 
theoretical justifications for this assumption. In the language of jigsaw puzzles,
is it the right shape, and does it have the right size, in the language of particle
physics, does it have the right spin, parity and couplings? We then discuss what
physics may lie behind and beyond it, and review how to probe these ideas with
possible future accelerators.

\subsection{What are its spin and parity?}

Since the $H(125)$ particle decays into pairs of photons, Lorentz invariance
assures us that it cannot have spin 1, but it might {\it a priori} have spin 0, 2
or higher and, in each case, it might have either positive or negative parity. 
Many tests of the $H(125)$ spin and parity have been proposed theoretically
and carried out by the LHC experiments. Examples include the polar angle distribution
in $H \to \gamma \gamma$ decays and final-state angular
correlations in $H \to W W^* \to \ell+ \ell^- +$ missing transverse energy decays
and in $H \to Z Z^* \to 2\ell+2 \ell^-$ decays~\cite{Hspin}. Also, the kinematics of $H(125)$
production mechanisms such as production in association with a $W$ or $Z$
boson would differ for different spin-parity assignments~\cite{EHSY}, which have been probed
by the Fermilab Tevatron experiments~\cite{Tevspin}.

One example of a spin-parity analysis of the $H(125)$ 
in the $X \to ZZ \to 2\ell^+ 2\ell^1$ final state is shown in Fig.~\ref{fig:Hspin}~\cite{CMSspin}.
This and all the other published analyses are in excellent agreement with the $J^P = 0^+$
spin-parity assignment predicted for the Higgs boson, and all the alternative
spin-parity assignments studied have been strongly excluded. These include
the pseudoscalar $0^-$ possibility and various spin-2 possibilities. The $H(125)$
passed these first important experimental tests with flying colours.

\begin{figure}[ht]
\begin{center}
\includegraphics[width=12cm]{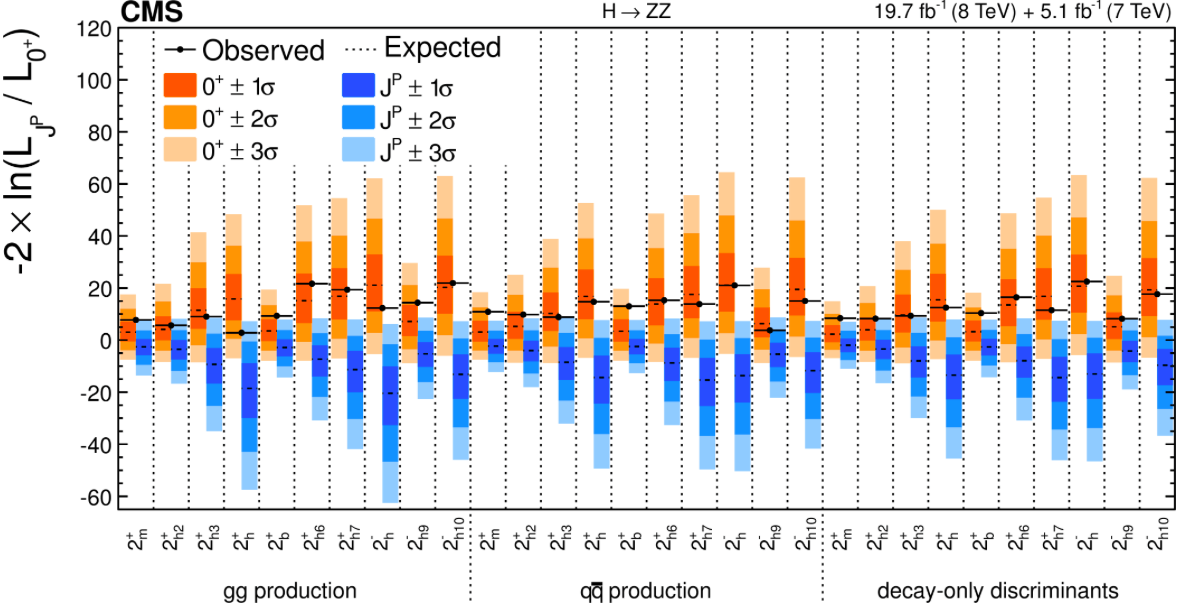}
\caption{\it Comparison between the Standard Model Higgs boson hypothesis in the
$X \to ZZ \to 2\ell^+ 2\ell^-$ final state with various spin-two $J^P$ hypotheses~\protect\cite{CMSspin}.}
\label{fig:Hspin}
\end{center}
\end{figure}

\subsection{Does it couple to particle masses?}

It is a fundamental property of the Higgs boson that, since the field vev
gives masses to the other elementary particles, the $H$ couplings to them
should be proportional to their masses, see (\ref{VVH}) and (\ref{fH}). One
way to test this is to analyze the $H(125)$ production and decay data
assuming couplings to other particles that are proportional to some 
nonlinear power of their masses~\cite{EY}:
\begin{equation}
y_f \; = \; \left( \frac{m_f}{M} \right)^{1 + \epsilon}, \quad g_V \; = \; \left( \frac{m_V^{2(1 + \epsilon)}}{M^{1 + 2 \epsilon}} \right) \, ,
\label{epsilon}
\end{equation}
where the unknown power $\epsilon$ and mass scale $M$ are to be fitted to
the data, the Standard Model expectations being $\epsilon = 0$ and $M = v = 246$~GeV.
The result of such an analysis performed jointly by the ATLAS and CMS
Collaborations is shown in Fig.~\ref{fig:Mepsilon}~\cite{ATLAS+CMScouplings}, including the best fit and the
68 and 95\% CL bands. The joint analysis yields
\begin{equation}
\epsilon = 0.023^{+0.029}_{-0.027}\, , \quad M \; = \; 233 \pm 13 \; {\rm GeV} \, ,
\label{ATLASCMSfit}
\end{equation}
which are highly compatible with the Standard Model predictions. In this way,
the $H(125)$ passed another crucial experimental test. We can also see explicitly
in Fig.~\ref{fig:Mepsilon} that the decay rates for $H \to \mu^+ \mu^-$ and $\tau^+ \tau^-$
must be very different, a first strong violation of lepton universality, as expected in the
mass-dependent couplings of the Higgs boson.

\begin{figure}[ht]
\begin{center}
\includegraphics[width=9cm]{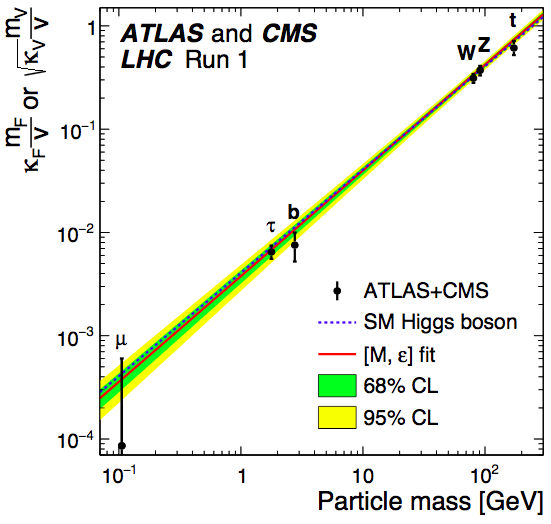}
\caption{\it A fit to a parametrization of the form (\protect\ref{epsilon}) to Higgs coupling
measurements by the ATLAS and CMS Collaborations in various channels.
The dotted line connects the Standard Model predictions, the best fit is shown as a
red line, and the 68 and 95\% CL ranges are shown as green and yellow bands~\protect\cite{ATLAS+CMScouplings}.}
\label{fig:Mepsilon}
\end{center}
\end{figure}

\subsection{Flavour-changing couplings?}

Flavour-changing couplings of the Higgs boson are expected to be very
suppressed in the Standard Model, though they might be present in
extensions with multiple Higgs multiplets. Upper limits on flavour-changing
interactions at low energies and dipole moments can be used to 
constrain the the possible flavour-changing interactions of the $H(125)$~\cite{BEI}.
Examples of relevant tree and loop diagrams involving $H$ are shown in Fig.~\ref{fig:BEI}.
Upper limits on flavour-changing quark interactions exclude the observability 
of quark-flavour-violating $H(125)$ decays, but lepton-flavour-violating decays
could be relatively large. We found that the branching ratio for either
$H \to \tau \mu$ or $\tau e$ (but not both) could be ${\cal O}(10)$\%,
comparable to the Standard Model prediction for BR$(H \to \tau^+ \tau^-)$,
whereas the branching ratio for $H \to \mu e$ must be $< 2 \times 10^{-5}$.
Analyses of LHC Run~1 data yielded results~\cite{CMSLFV,ATLASLFV} compatible with these upper limits:
\begin{eqnarray}
{\rm CMS}: \; {\rm BR}(H \to \tau \mu) & = & 0.84^{+0.39}_{-0.37} \%, \; 
{\rm BR}(H \to \tau e) \; < \; 0.69 \%, \; {\rm BR}(H \to e \mu) \; < \; 0.036 \% \, , \nonumber \\
{\rm ATLAS}: \; {\rm BR}(H \to \tau \mu) & = & 0.77 \pm 0.62 \% \, .
\label{taumu}
\end{eqnarray}
That said, the CMS result for BR$(H \to \tau \mu)$, in particular, whetted theoretical appetites
for additional data from Run~2 of the LHC. A first preliminary result from CMS does not
indicate any deviation from the Standard Model~\cite{CMS2}, but this is definitely {\it une affaire \`a suivre}!

\begin{figure}[t]
\begin{center}
\includegraphics[width=3.25cm]{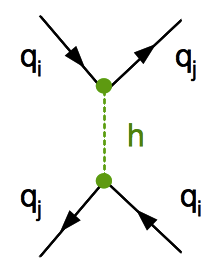}
\hskip 1 cm
\includegraphics[width=7.5cm]{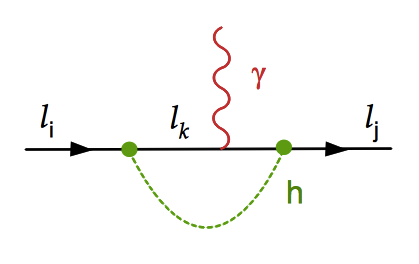}
\vskip -1 cm
\end{center}
\caption{\label{fig:BEI} \it Left panel: Tree-level $H$-exchange
diagram that may contribute to a generic flavour-changing amplitude.
Right panel: One-loop $H$-exchange diagram that may contribute to anomalous magnetic and electric dipole
moments of charged leptons ($i=j$), or to radiative lepton-flavour-violating decays  ($i \ne j$)~\protect\cite{BEI}. }
\end{figure}

\subsection{Loop-induced couplings}

As mentioned in the first Lecture, two of the most important Higgs couplings
are induced by loop diagrams, namely the $gg H$ vertex responsible for the
dominant $H$ production mechanism, which is mainly generated by the top
quark in the Standard Model, and the $H \gamma \gamma$ vertex responsible
for one of the most distinctive $H(125)$ decays, which is mainly generated
by loops of top quarks and $W^\pm$ bosons in the Standard Model, as shown in Fig.~\ref{fig:LOgg}. Via
these vertices, the Run~1 LHC data have already provided important
consistency checks on the Standard Model predictions for the $H$ couplings
at the quantum level.
Fig.~\ref{fig:gluonphoton} displays the combined ATLAS and CMS constraints
on the magnitudes of the $H \gamma \gamma$ and $gg H$ couplings relative
to their Standard Model values~\cite{ATLAS+CMScouplings}. We see good consistency at the 10 to 20\%
level, which also provides significant restrictions on possible extensions
of the Standard Model such as a fourth generation, supersymmetric particles
and heavy vector-like quarks.

\begin{figure}[ht]
\begin{center}
\includegraphics[width=9cm]{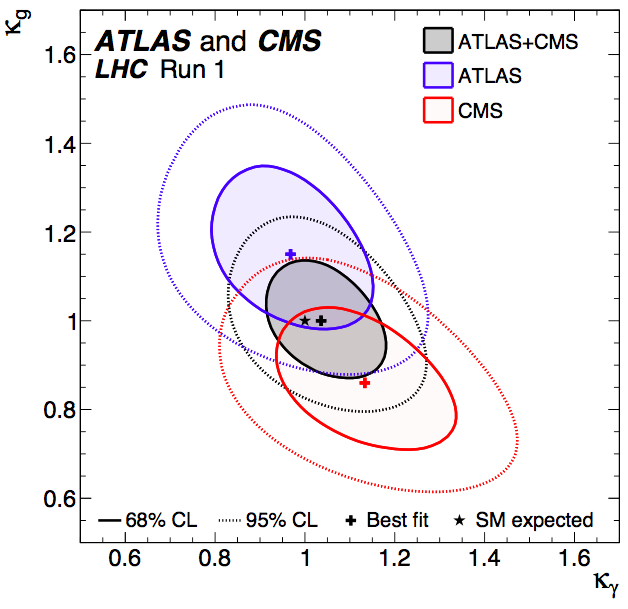}
\caption{\it A fit by the ATLAS and CMS Collaborations to the magnitudes of the 
$H \gamma \gamma$ and $gg H$ couplings, normalized by factors $(\kappa_\gamma, \kappa_g)$ relative
to their Standard Model values~\protect\cite{ATLAS+CMScouplings}.}
\label{fig:gluonphoton}
\end{center}
\end{figure}

\subsection{Is it elementary or composite?}

Broadly speaking, there are two schools of theoretical thought about this question.

On the one hand, many theorists are attracted by the idea of an elementary Higgs
scalar field, as in the original formulation, but are concerned by the problems
connected with loop corrections to the Higgs mass. Quantum corrections to the 
mass parameter $\mu$ in the effective potential (\ref{Vphi}) due to, e.g., the top
quark or the Higgs self-coupling, exhibit quadratic
divergences. If one cuts the loop integrals off at some momentum scale $\Lambda$,
one is left with large residual contributions if $\Lambda$ is identified with some
high new physics scale such as that of grand unification or the Planck mass.
In the case of a loop of fermions $f$ such as the top quark, shown in Fig.~\ref{fig:higgscorr1}(a), one finds 
\begin{equation}
\Delta m_H^2 \; = \; -\frac{y_f^2}{8\pi^2}[2\Lambda^2 + 6m_f^2 \ln(\Lambda/m_f)+...] \, ,
\label{quadf}
\end{equation}
where $y_f$ is the Yukawa coupling and the $\dots$ represent non-divergent mass-dependent terms, 
and in the case of a loop of scalars $S$, shown in Fig.~\ref{fig:higgscorr1}(b),
one finds similar divergent contributions:
\begin{equation}
\Delta m_H^2 \; = \; \frac{\lambda_S}{16\pi^2}[\Lambda^2 - 2m_S^2 \ln(\Lambda/m_S)+...] \, .
\label{quadS}
\end{equation}
If the Standard Model
were to remain valid up to the Planck scale, $M_P\simeq 10^{19}$ GeV, so that $\Lambda=M_P$,
each of the quadratic ``corrections" would be $\simeq 10^{34}$ times larger 
than the physical mass-squared of the Higgs, namely ${\cal}(10^4)$~GeV$^2$.

\begin{figure}
\vspace{-26cm}
\begin{center}
\includegraphics[height=35cm]{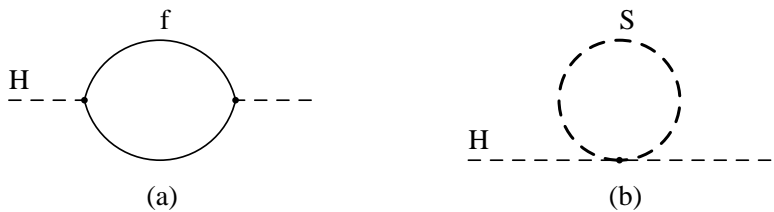}
\vspace{-6cm}
\caption{\it One-loop quantum corrections to the mass-squared of the Higgs boson due to (a) 
the loop of a generic fermion $f$, (b) a generic scalar $S$.
\label{fig:higgscorr1}}
\end{center}
\end{figure}

The relatively small physical value of the Higgs mass is not protected by any
symmetry of the Standard Model, and keeping it small seems to require some
unnatural fine-tuning unless there is some suitable new physics at the TeV scale.
The favoured example of such new physics in an elementary Higgs scenario is
supersymmetry, which exploits the opposite signs of loop corrections due to
fermions and bosons in (\ref{quadf}, \ref{quadS}). If these occur in pairs with related couplings:
\begin{equation}
\lambda_S \; = \; 2 \lambda_f^2
\label{lambdarelation}
\end{equation}
as in supersymmetric models~\cite{hierarchy}, and if the differences between the masses of
supersymmetric partners are ${\cal O}(1)$~TeV so that the subleading terms in (\ref{quadf}, \ref{quadS})
are not large, the quadratic term $\mu$ in the Higgs potential (\ref{Vphi}) is kept naturally small,
and hence also the Higgs mass and the electroweak scale.
One of the miracles of supersymmetry is that the symmetry between fermions and bosons
cancels not only the one-loop quadratic
divergences (\ref{quadf}, \ref{quadS}), but also all quadratic divergences
at higher order in perturbation theory, as well as many logarithmic divergences~\cite{WZ}.

On the other hand, many other theorists believe that the Higgs is
composite, a bound state of fermions like Cooper pairs in BCS
superconductivity and pions in QCD. Such a theory has a natural cut-off
at the scale of the strongly-interacting composite dynamics, analogous to
$\Lambda_{QCD}$. The Standard Model does not contain any candidate
for this new strong dynamics, and attention focused initially on some
scaled-up version of QCD~\cite{technicolour}. However, simple models of this type were
incompatible with the precision electroweak data mentioned in the first
Lecture, and predicted a heavy strongly-interacting scalar particle unlike
the Higgs boson discovered in 2012. Accordingly, attention has shifted to
an alternative idea that the Higgs boson is analogous to the pion of QCD,
namely that it is a pseudo-Nambu-Goldstone boson of some larger chiral
symmetry that is broken down to the Standard Model, much like the pion in QCD~\cite{littleH}. In such a model,
the lightness of the Higgs boson is enforced by this approximate chiral symmetry.
Generic features of such theories include a coloured top partner fermion
that cancels the one-loop Higgs mass corrections due to the top quark,
some new scalars and/or gauge bosons with relatively low masses 
$\lesssim 1$~TeV, and a strongly-interacting ultraviolet completion at a mass scale
that is ${\cal O}(10)$~TeV.

A convenient way to parametrize the phenomenology of such a theory is
to assume that the Higgs sector has an underlying SU(2)$\times$SU(2)
structure that is broken down to a custodial SU(2) symmetry so as to
retain the successful tree-level relation $\rho \equiv m_W/m_Z \cos \theta_W \simeq 1$.
The Goldstone bosons $\pi_a: a = 1, 2, 3$ of this symmetry-breaking pattern that are
`eaten' by the $W^\pm$ and $Z$ to become their longitudinal polarization states are then parametrized
by a traceless $2\times2$ matrix $\Sigma = \exp (i \sigma_a \pi_a/v )$, with the following couplings
to the Higgs boson $H$:
\begin{eqnarray}
{\cal L} & = & \frac{v^2}{4} {\rm Tr} D_\mu \Sigma D^\mu \Sigma \left(1 + 2 a \frac{H}{v} + b \frac{H^2}{v^2} + \dots \right)
- m_i \bar{\psi}^i_L \Sigma \left( c \frac{H}{v} + \dots \right) + {\rm h.c.} \nonumber \\
&& + \frac{1}{2} \partial_\mu H \partial^\mu H + \frac{1}{2} H^2 + d_3 \frac{1}{6} \left( \frac{3 m_H^2}{v} \right) H^3
+ d_4 \frac{1}{24} \left( \frac{3 m_H^2}{v} \right) H^4 + ... \, ,
\label{nonlinearH}
\end{eqnarray}
where the coefficients $a, b, c, \dots$ are normalized so that they are all unity in the Standard Model.
The task of experiments is then to measure these coefficients and see whether they differ from
these predictions, as they may in composite Higgs models. For example, in two minimal composite 
Higgs models MCHM$_{4,5}$ one has
\begin{eqnarray}
{\rm MCHM}_4 & : & a \; = \; \sqrt{1 - \zeta}, \quad c \; = \; \sqrt{1 - \zeta} \, , \nonumber \\
{\rm MCHM}_5 & : & a \; = \; \sqrt{1 - \zeta}, \quad c \; = \; \frac{1 - 2 \zeta}{\sqrt{1 - \zeta}} \, ,
\label{MCHMs}
\end{eqnarray}
where $\zeta$ is a model parameter that is not specified {\it a priori}.

Fig.~\ref{fig:kappas} shows two experimental analyses of the data, using a different
notation: $a \to \kappa_V, c \to \kappa_F$. The left panel compares the constraints
from Higgs data alone (yellow and orange ellipses) with the result of a global analysis
including also precision electroweak data (blue ellipses)~\cite{Gfitterkappas}. We see that these are
largely complementary, with the Higgs data constraining $\kappa_F = c$ and the
electroweak data constraining $\kappa_V = a$. The Standard Model prediction
$\kappa_F = c = 1$, $\kappa_V = a = 1$ is close to the best-fit point and well within
the global 68\% CL contour. The right panel compares ATLAS and CMS Higgs
measurements with the predictions (\ref{MCHMs}) of the MCHM$_{4, 5}$ models~\cite{oldATLAS+CMS}.
We see that the data require $\zeta \lesssim 0.1$, necessitating some tuning of
these models so that their predictions resemble those of the Standard Model.

\begin{figure}[ht]
\begin{center}
\includegraphics[width=7.3cm]{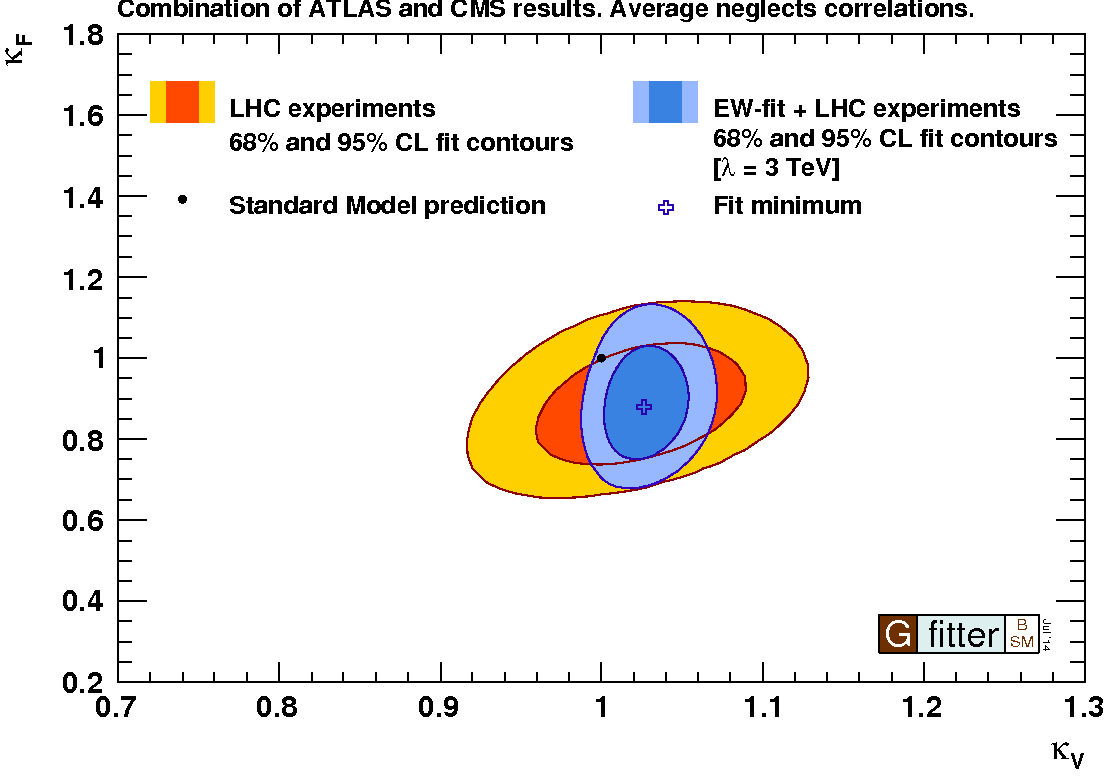}
\includegraphics[width=7.7cm]{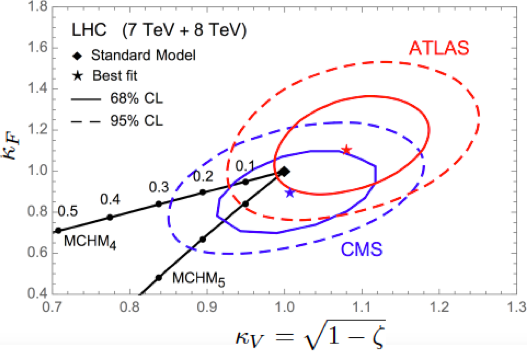}
\caption{\it Left panel: A fit by the Gfitter Group to the LHC $H$ coupling
measurements (orange and yellow ellipses) and in combination with precision
electroweak data (blue ellipses)~\protect\cite{Gfitterkappas}. Right panel: Comparison of ATLAS and CMS constraints
on $H$ couplings with predictions of the MCHM$_{4,5}$ models~\protect\cite{oldATLAS+CMS}.}
\label{fig:kappas}
\end{center}
\end{figure}

Fig.~\ref{fig:KappaInterplay} shows how the different Higgs coupling measurements by
ATLAS and CMS combine to give their overall constraints on $(\kappa_V, \kappa_F)$~\cite{ATLAS+CMScouplings}.
Most of the measurements are relatively insensitive to the sign of $\kappa_F$,
the exception being that of the $H \gamma \gamma$ coupling. Its sensitivity is
due to the interference between the top and $W^\pm$ loop contributions to the
coupling, which interfere destructively for the Standard Model (positive) sign of
$\kappa_F$ and constructively for the non-standard (negative) sign. Largely as
a result of this asymmetry, the combined fit decisively favours the Standard Model
sign of $\kappa_F$. Measuring the cross section for single $t/{\bar t} H$ production could also
provide a direct determination of this sign, and could probe the possible existence
of a CP-violating $t{\bar t} H$ coupling~\cite{EHST}.

\begin{figure}[ht]
\begin{center}
\includegraphics[width=9cm]{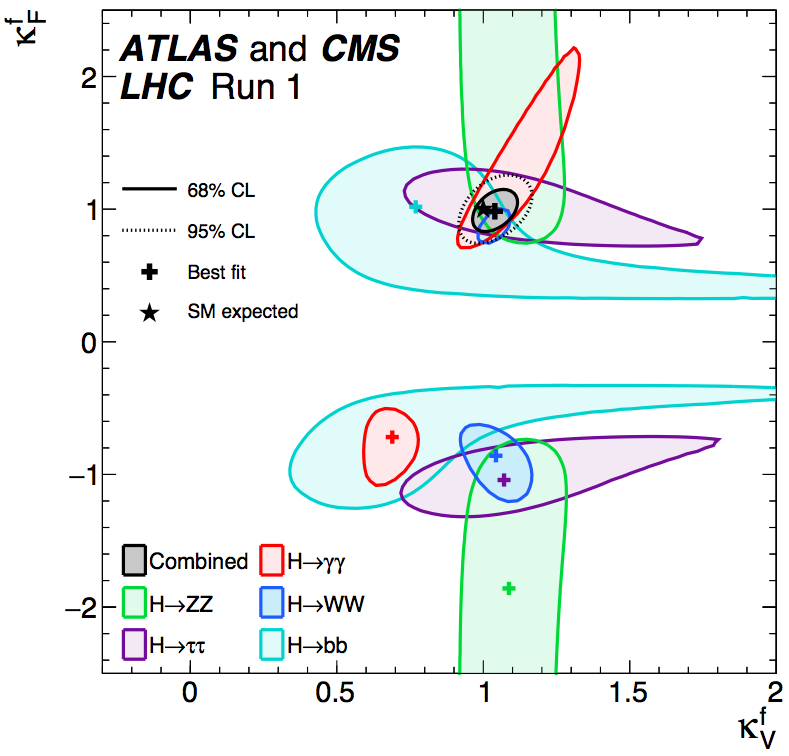}
\caption{\it A fit by the ATLAS and CMS Collaborations to the magnitudes of the 
$HVV$ and $H {\bar f}f$ couplings, normalized by factors $(\kappa_V, \kappa_F)$ relative
to their Standard Model values~\protect\cite{ATLAS+CMScouplings}.}
\label{fig:KappaInterplay}
\end{center}
\end{figure}

The general conclusion of these analyses using the parametrization (\ref{nonlinearH})
is that there is no indication of any deviation from the Standard Model predictions for
the $H$ couplings of the form that might have arisen in a composite Higgs scenario.
Introducing a single global modification factor $\mu$ for the $H$ couplings to Standard
Model particles, the combined ATLAS and CMS data imply that~\cite{ATLAS+CMScouplings}
\begin{equation}
\mu \; = 1.09^{+0.11}_{-0.10} \; = \; 1.09 \pm 0.07 \, ({\rm stat.}) \pm 0.04 \, ({\rm expt}) \pm 0.03 \, ({\rm thbgrd}) ^{+0.07}_{-0.06} \, ({\rm thsig}) \, ,
\label{overallmu}
\end{equation}
where the last three uncertainties are systematics. Thus, overall the strength of the Higgs
couplings agrees with the Standard Model at the $\sim 10$\% level, though individual
couplings have larger uncertainties. Moreover, there is no evidence for any
decays of $H$ to unknown particles, in particular invisible decays. The $H(125)$ particle
looks very much the way it was predicted in the Standard Model.

For this reason, the Physics Class of the Swedish Academy stated in its citation for the
2013 Nobel Prize~\cite{NP} that {\it Today we believe that ``Beyond any reasonable doubt, it is a Higgs 
boson."}~\footnote{This quotation was taken from the preprint version of~\cite{EY}. They
apparently did not notice that this phrase was removed from the published version of~\cite{EY}
at the insistence of the anonymous referee, who considered that ``Beyond any reasonable doubt"
is not a scientific statement.}.

\subsection{The Standard Model effective field theory}

At this point a popular approach is to assume that the $H(125)$ particle is exactly
Standard Model-like, and use it in a model-independent search for new physics that
could manifest itself via higher-dimensional effective interactions between Standard Model fields,
in particular those of dimension 6~\cite{dim6}:
\begin{equation}
{\cal L}_{eff} \; = \; \sum_n \frac{c_n}{\Lambda^2} {\cal O}_n \, ,
\label{dim6}
\end{equation}
where $\Lambda$ is some characteristic scale of new physics and the $c_n$ are unknown
dimensionless coefficients. These can be constrained by a combination of data on Higgs
properties, precision electroweak data, triple-gauge couplings (TGCs), etc.. The beauty of
this approach is that it provides an integrated framework for analyzing all these categories
of data in a unified and consistent way.

This attractive approach is, however, unwieldy when applied in full generality, because
of the large number of possible dimension-6 operators, even if one assumes the
SU(2)$\times$U(1) symmetry of the Standard Model. For this reason, one often makes
simplifying assumptions, e.g., about the flavour structure of the operators. Furthermore,
if one restricts attention to precision electroweak observables, Higgs and TGC
measurements, global fits to these data become manageable. Table~3 lists the
CP-even dimension-6 operators~\cite{PomarolRiva} relevant for these measurements. In each case,
we also indicate the categories of observables that provide the greatest sensitivities to the operator coefficients.

\vspace{0.5cm}
\begin{table}[h]
{\small
\begin{center}
\begin{tabular}
{ | c | c | c | } 
\hline
EWPTs & Higgs Physics & TGCs \\
\hline
\multicolumn{3}{| c |}{ ${\mathcal O}_W=\frac{ig}{2}\left( H^\dagger  \sigma^a \leftrightarrow {D^\mu} H \right)D^\nu  W_{\mu \nu}^a$  } \\
\hline
\multicolumn{2}{| c |}{ ${\mathcal O}_B=\frac{ig'}{2}\left( H^\dagger  \leftrightarrow {D^\mu} H \right)\partial^\nu  B_{\mu \nu}$  } & ${\mathcal O}_{3W}= g \frac{\epsilon_{abc}}{3!} W^{a\, \nu}_{\mu}W^{b}_{\nu\rho}W^{c\, \rho\mu}$	\\
\hline
${\cal O}_T=\frac{1}{2}\left (H^\dagger {\leftrightarrow{D}_\mu} H\right)^2$ & \multicolumn{2}{| c |}{ ${\mathcal O}_{HW}=i g(D^\mu H)^\dagger\sigma^a(D^\nu H)W^a_{\mu\nu}$ } \\
\hline
$\mathcal{O}_{LL}^{(3)\, l}=( \bar L_L \sigma^a\gamma^\mu L_L)\, (\bar L_L \sigma^a\gamma_\mu L_L)$ & \multicolumn{2}{| c |}{$ {\mathcal O}_{HB}=i g^\prime(D^\mu H)^\dagger(D^\nu H)B_{\mu\nu}$ } \\
\hline
${\mathcal O}_R^e = (i H^\dagger {\leftrightarrow { D_\mu}} H)( \bar e_R\gamma^\mu e_R)$ & ${\mathcal O}_{g}=g_s^2 |H|^2 G_{\mu\nu}^A G^{A\mu\nu}$ &  \\
\hline
${\cal O}_{R}^u = (i H^\dagger {\leftrightarrow { D_\mu}} H)( \bar u_R\gamma^\mu u_R)$ & ${\mathcal O}_{\gamma}={g}^{\prime 2} |H|^2 B_{\mu\nu}B^{\mu\nu}$ & \\
\hline
${\cal O}_{R}^d = (i H^\dagger {\leftrightarrow { D_\mu}} H)( \bar d_R\gamma^\mu d_R)$ & ${\mathcal O}_H=\frac{1}{2}(\partial^\mu |H|^2)^2$ & \\
\hline
${\cal O}_{L}^{(3)\, q}=(i H^\dagger \sigma^a {\leftrightarrow { D_\mu}} H)( \bar Q_L\sigma^a\gamma^\mu Q_L)$  & ${\mathcal O}_{f}   =y_f |H|^2    \bar{F}_L H^{(c)} f_R + \text{h.c.}$ & \\
\hline
${\cal O}_{L}^q=(i H^\dagger {\leftrightarrow { D_\mu}} H)( \bar Q_L\gamma^\mu Q_L)$  & $\mathcal{O}_6 = \lambda|H|^6$ & \\ 
\hline
\end{tabular}
\end{center}
}
\caption{\it The relevant CP-even dimension-6 operators in the basis~\protect\cite{PomarolRiva} that we use. For each operator,
we list the categories of observables that provide the greatest sensitivities to the operator.}
\label{tab:dim6}
\end{table}

The left panel of Fig.~\ref{fig:SMEFT1} shows results from a fit to precision data on
leptonic electroweak observables~\cite{ESY}.
The lower horizontal axis shows the
possible numerical values of these coefficients, and the upper horizontal axis shows the
corresponding new physics scales. Here and in subsequent plots, the green bars are for
fits to individual operator coefficients assuming the other operators are absent, and the
bars of other colours are for global fits marginalizing of all the operators that could
contribute. In general, these coloured bars extend further than the green bars. The right
panel of Fig.~\ref{fig:SMEFT1} extends this analysis to include hadronic electroweak
observables~\cite{ESY}. The vertical dashed lines in the two panels are for the same new physics scale,
and serve to emphasize the point that the constraints on hadronic observables are, in
general, weaker than those on leptonic observables and, moreover, some exhibit deviations from
the Standard Model predictions that remain to be understood. However, in both cases the
new-physics constraints are in the multi-TeV range.

\begin{figure}[ht]
\begin{center}
\includegraphics[width=15cm]{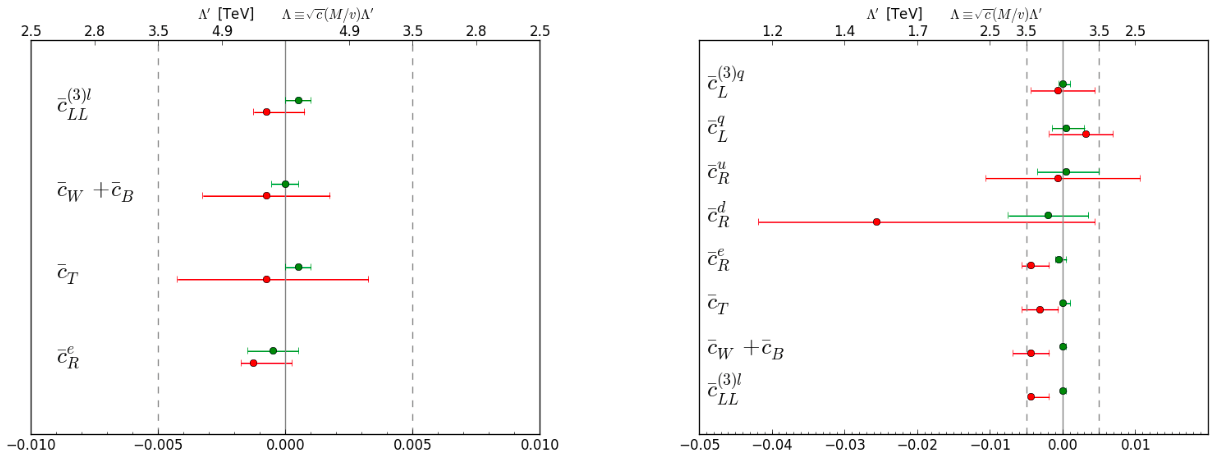}
\caption{\it The 95\% CL ranges from an analysis~\protect\cite{ESY} of precision leptonic electroweak observables (left panel)
and including also hadronic electroweak observables (right panel). The upper (green) bars 
denote fits to individual operator coefficients, and the lower (red) bars are for
marginalized multi-operator fits. The upper axis should be read with factor
$m_W/v \sim 1/3$ for the combination $\bar{c}_W + \bar{c}_B$.}
\label{fig:SMEFT1}
\end{center}
\end{figure}

The left panel of Fig.~\ref{fig:SMEFT2} shows results from global fits to data on 
Higgs production strengths and kinematics (blue bars), to data on TGCs (red bars),
and their combination (black bars)~\cite{ESY}. The green bars again show the results of fits
to individual operator coefficients, which are generally smaller than the other bars. 
We note that the constraints on the new-physics
scale from Higgs and TGC data shown in the left panel of Fig.~\ref{fig:SMEFT2} are,
in general, weaker than from the precision electroweak data: they are typically only
a fraction of a TeV. The right panel Fig.~\ref{fig:SMEFT2} emphasizes the complementarity of Higgs and
TGC measurements, as reflected in anomalous TGC couplings. The orange and yellow
ellipses show the constraints from direct TGC measurements, whereas the green
ellipses show the indirect constraints from Higgs measurements, and the blue ellipses
show the results of a global fit~\cite{Falkowski}.

\begin{figure}[ht]
\begin{center}
\includegraphics[width=8cm]{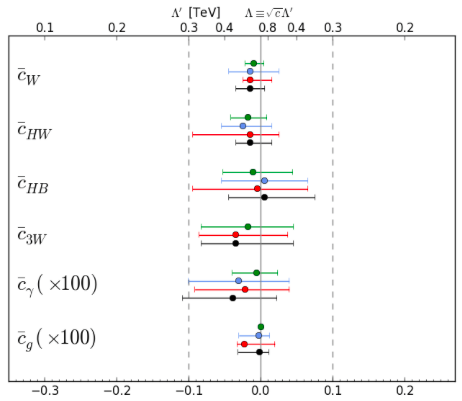}
\includegraphics[width=7cm]{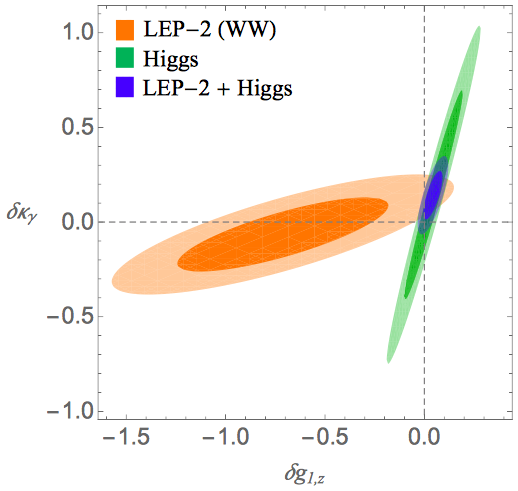}
\caption{\it Left panel: The 95\% CL ranges for individual operator fits (green bars), 
and the marginalised 95\% ranges for multi-operator fits. The blue bars
combine the LHC signal-strength data with the kinematic distributions for associated 
$H+V$ production measured by the ATLAS and D0 Collaborations, the
red bars includeh the LHC TGC data, and the black bars show results from a
global combination with both the associated production and TGC data~\protect\cite{ESY}. 
Note that the coefficients $\bar{c}_{\gamma,g}$ are shown magnified by factors
of 100, so for these coefficients the upper axis should be read with a factor of 10.
Right panel: The 68 and 95\% CL ranges in the plane of anomalous TGCs $(\delta g_{1,z}, \delta \kappa_\gamma)$
including LEP TGC constraints, LHC Higgs data and their combination~\protect\cite{Falkowski}.}
\label{fig:SMEFT2}
\end{center}
\end{figure}

The Standard Model effective field theory is the preferred framework for analyzing
future LHC data. Fig.~\ref{fig:LHCEFT} shows the results of global fits to
LHC Higgs data using only production rates (left panel) and including production
kinematics (right panel)~\cite{LHCHXSWG4}. In each case, the blue bars are obtained from an analysis
of present data, the green bars illustrate the prospective sensitivities with 300/fb
of data, and the red bars those with 3000/fb of data. The prospective sensitivities
are impressive, particularly when the kinematical information is included.

\begin{figure}[ht]
\begin{center}
\includegraphics[width=10cm]{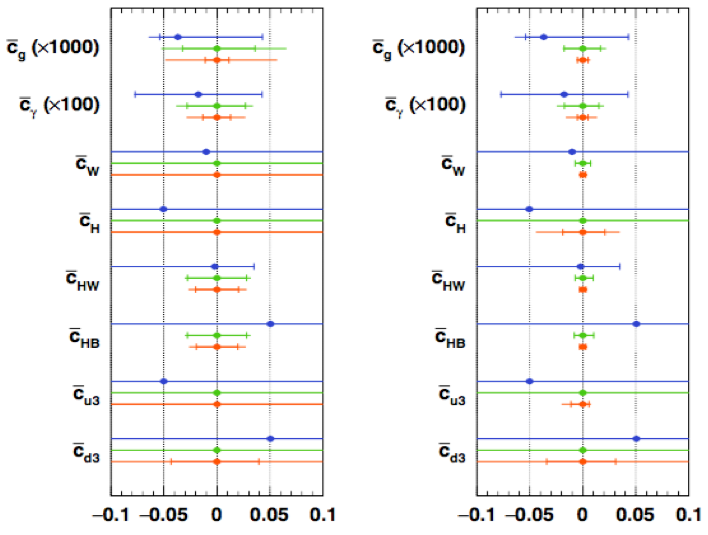}
\caption{\it Results of present and prospective global fits to
LHC Higgs data using only production rates (left panel) and including production
kinematics (right panel). The blue bars are obtained from an analysis
of present data, the green bars illustrate the prospective sensitivities with 300/fb
of data, and the red bars those with 3000/fb of data~\protect\cite{LHCHXSWG4}.}
\label{fig:LHCEFT}
\end{center}
\end{figure}

\subsection{Beware of historical hubris}

Despite the continuing absence of any direct evidence for the new physics beyond the
Standard Model at the LHC, one should not become disheartened. History abounds
with examples of people who thought they knew it all, but did not. In 1894, just before
the discoveries of radioactivity and the electron, Albert Michelson declared that
``The more important fundamental laws and facts of physical science have all been discovered"~\cite{Michelson}.
More recently, prior to the
string revolution, Stephen Hawking asked ``Is the End in Sight for Theoretical Physics?"~\cite{Hawking}.
However, my favourite example of a lack of ability to think outside the box is the Spanish 
Royal Commission that rejected a proposal by Christopher Columbus to sail west
before 1492: ``So many centuries after the Creation, it is unlikely that anyone could 
find hitherto unknown lands of any value"~\cite{Columbus}. Many of us have seen referees' reports
with a similar flavour.

\subsection{The Standard Model is not enough}

The title of this Subsection is a paraphrase of the title of a James Bond movie~\cite{Bond} and,
in deference to him, one may cite 007 reasons for anticipating physics beyond the
Standard Model. 001) As discussed in Lecture~1, within the Standard Model the electroweak 
vacuum is unstable against decay to high $H$ field values. 002) The Standard Model
has no candidate for the astrophysical dark matter. 003) The Cabibbo-Kobayashi-Maskawa (CKM)
Model does not explain the origin of the matter in the universe. 004) The Standard Model does not have
a satisfactory mechanism for generating neutrino masses. 005) The Standard Model
does not explain or stabilize the hierarchy of mass scales in physics. 006) The Standard
Model does not have a satisfactory mechanism for cosmological inflation. 007) We need
a quantum theory of gravity.

Several of these issues will be addressed by LHC measurements during Run 2, e.g.,
the top quark mass will be determined more accurately, there will be searches for
dark matter particles, there will be searches for CP violation and other flavour physics
beyond the CKM model, as well as new particles that could help stabilize the electroweak
scale. Personally, I am a fan of supersymmetry as a framework that could solve or at
least mitigate many of the problems on James Bond's list, so I focus now on that theory.

\subsection{Supersymmetry}

Supersymmetry is an extension of the Standard Model that has long been favoured by
many theorists~\cite{susy}. Some are disappointed that it has not yet appeared at the LHC, but
then neither has any other proposed extension of the Standard Model such as compositeness
or extra dimensions. Rather, I would argue that Run~1 of the LHC has provided three
new additional reasons to favour supersymmetry.

One is the apparent instability of the electroweak vacuum within the Standard Model,
which can be stabilized by a theory resembling supersymmetry, as discussed in Subsection 1.10. 
Specifically, in a supersymmetric theory the negative running of the Higgs quartic 
self-coupling $\lambda$ due to the top quark loop is exactly cancelled by stop squark loops.
Moreover, the negative sign of the quadratic term in the Higgs potential (\ref{Vphi}) and
hence the appearance of the electroweak vacuum can be understood dynamically as a
different effect of renormalization by the heavy top quark via the logarithmic terms in (\ref{quadf} \ref{quadS}).

A second Run~1 motivation for supersymmetry is the mass of the Higgs boson. Minimal
supersymmetric models predicted that it should weigh $\lesssim 130$~GeV~\cite{EFZ}, as discussed
in more detail in the next Subsection, in agreement
with the measurement (\ref{measuredmH}). This is because supersymmetry actually
predicts the magnitude of the Higgs quartic self-coupling: $\lambda \sim g^2 + g^{\prime 2}$,
where $g$ and $g^\prime$ are the SU(2) and U(1) couplings of the Standard Model.

The third Run-1 motivation for supersymmetry is that simple supersymmetric models
also predicted that the Higgs couplings should be with a few \% of the Standard Model
values, in perfect consistency with the measurements to date~\cite{EHOW}. In fact, in a supersymmetric
model it may be very difficult to measure any deviations from the Standard Model predictions
for the Higgs couplings~\cite{Interplay}, as discussed later.

These new reasons for liking supersymmetry are in addition to the many traditional
reasons to like it, such as its ability to stabilize the mass hierarchy via the
cancellation of the quadratic divergences in loop corrections to the Higgs mass (\ref{quadf} \ref{quadS})~\cite{hierarchy}, the fact that it
naturally predicts a cold dark matter particle~\cite{EHNOS}, the fact that it improves the accuracy
of the grand unification prediction for $\sin^2 \theta_W$~\cite{s2thetaW}, and its essential r\^ole in
the superstring framework for a theory of quantum gravity.

\subsection{Higgs bosons in supersymmetry}

Even the minimal supersymmetric extension of the Standard Model (MSSM) requires
two complex Higgs doublets, in order to cancel out anomalous triangle diagrams with
higgsinos that would otherwise destroy the renomalizability of the theory, and in order
to give masses to all the quarks. These two complex Higgs doublets contain 8
degrees of freedom, of which 3 are combined with the massless $W^\pm$ and $Z$
fields to give them masses, as discussed in Subsection 1.4. There remain 5 degrees
of freedom that manifest themselves as massive Higgs bosons, two neutral scalars $h, H$,
one neutral pseudoscalar $A$ and two charged bosons $H^\pm$.

The tree-level Higgs mass-squared matrix has the following form in the MSSM:
\begin{equation}
\mathcal{M}{^{2,N=1}_{\text{tree}}}=
\begin{pmatrix}
m_Z^2\cos^2\beta+m_A^2\sin^2\beta & -(m^2_A+m^2_Z)\cos\beta\sin\beta \\
-(m^2_A+m^2_Z)\cos\beta\sin\beta & m_Z^2\sin^2\beta+m_A^2\cos^2\beta
\end{pmatrix} \, .
\label{massmatrixN=1}
\end{equation}
Diagonalizing (\ref{massmatrixN=1}), we find that the masses of the two scalars at the classical (tree) level can be written as
\begin{equation}
m^2_{h, H} \; = \; \frac{1}{2} \left(m_A^2 + m_Z^2 \mp \sqrt{(m_A^2 + m_Z^2)^2 - 4 m_Z^2 m_A^2 \cos^2 2 \beta} \right) \, ,
\label{treemasses}
\end{equation}
where $\tan \beta$ is the ratio of the vevs of the 2 Higgs doublets. At face value,
the formula (\ref{treemasses}) implies that the lighter neutral scalar Higgs boson $h$
should have a mass $< m_Z$. However, there is an important one-loop correction to $m_h$
due to the stop squarks ${\tilde t}_{1,2}$:
\begin{equation}
\Delta m_h^2=\frac{3m_t^4}{4\pi^2 v^2}\ln\left(\frac{m_{\tilde{t}_1}m_{\tilde{t}_2}}{m_t^2}\right) 
+ \dots \, ,  
\label{loopcorrection}
\end{equation}
which can increase $m_h$ by $\lesssim 40$~GeV, as seen in Fig.~\ref{fig:susyHmasses}.
As also seen there, if the other Higgs bosons $H, A$ and $H^\pm$ are heavy, they are expected
to all be quite degenerate in mass. A curiosity of Fig.~\ref{fig:susyHmasses} is the
possibility that the {\it heavier} neutral scalar $H$ might weigh $\lesssim 130$~GeV,
with the $h$ even lighter~\cite{lighterh}. It may be difficult to reconcile this possibility with the LHC
measurements of the couplings of the $H(125)$, but the possibility of a lighter Higgs
boson should not be discounted completely, and further experimental searches in the
low-mass range are welcome!

\begin{figure}[ht]
\begin{center}
\includegraphics[width=9cm]{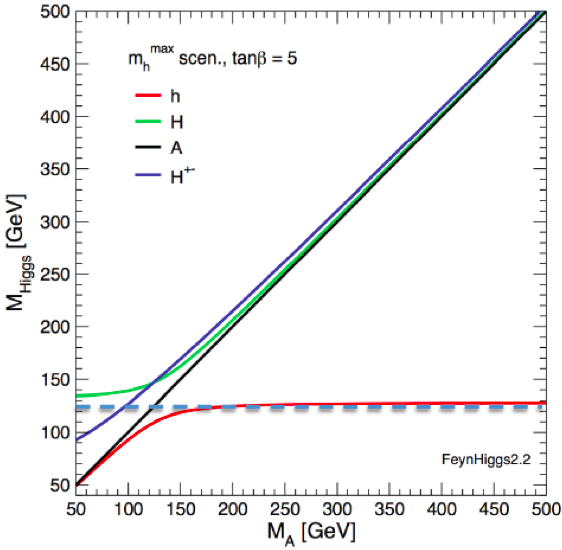}
\caption{\it A representative calculation of MSSM Higgs masses in the $m_h^{\rm max}$
scenario with $\tan \beta = 5$, using the {\tt FeynHiggs~2.2} code~\cite{FH}. Note a region at larger $m_A$
where the lighter neutral scalar Higgs boson may weigh $\sim 125$~GeV~\protect\cite{EFZ}, and a smaller region at
small $m_A$ where the heavier neutral scalar Higgs boson may weigh $\sim 125$~GeV~\protect\cite{lighterh}.}
\label{fig:susyHmasses}
\end{center}
\end{figure}

\subsection{More supersymmetry, not less?}

The Standard Model contains chiral fermions, i.e., the left- and right-handed fermion
states live in inequivalent representations of the SU(2)$\times$U(1) gauge group. As such,
they can be accommodated only within supermultiplets of simple $N = 1$ supersymmetry.
Theories with $N \ge 2$ supersymmetries would require left- and right-handed fermions
to transform identically under SU(2)$\times$U(1), so phenomenological supersymmetric
models such as the MSSM are usually restricted to $N = 1$. However, left-right symmetric
(vector-like) fermions appear in many extensions of the Standard Model, such as 
models with extra dimensions, string compactifications and some grand unified theories.
These extensions of the Standard Model could accommodate $N=2$ supersymmetry.

In the MSSM, although the quarks and leptons are chiral, the Higgs representations
forma vector-like pair, and so could in principle also accommodate $N = 2$ supersymmetry~\cite{EQS}.
One could consider the possibility that the Higgs sector is just the tip of an $N = 2$
supersymmetric iceberg, which would also include an $N=2$ gauge sector and possibly
vector-like fermion supermultiplets, as occurs in some string compactifications and grand
unified theories.

{\it So, what would an $N=2$ Higgs sector look like?}

In this case, differently from (\ref{loopcorrection}), the tree-level mass-squared matrix is:
\begin{equation}
\mathcal{M}{^{2,N=2}_{\text{tree}}}=
\begin{pmatrix}
m_Z^2\cos^2\beta+m_A^2\sin^2\beta & - (m^2_A-m^2_Z)\cos\beta\sin\beta \\
- (m^2_A-m^2_Z)\cos\beta\sin\beta & m_Z^2\sin^2\beta+m_A^2\cos^2\beta
\end{pmatrix} \, .
\label{massmatrixN=2}
\end{equation}
The crucial difference: is the replacement: $m^2_A+m^2_Z \to m^2_A-m^2_Z$ in the off-diagonal terms 
between the $N = 1$ and $N = 2$ cases (\ref{massmatrixN=1}) and (\ref{massmatrixN=2}).
The corresponding tree-level Higgs masses after diagonalization are shown in Fig.~\ref{fig:treemasses}.
In the $N = 1$ case (left panel) we see level repulsion for $m_A \sim 100$~GeV. However,
in the $N = 2$ case (right panel) we see linear level crossing with
\begin{equation}
m_h^{N=2} \; = \; m_Z, \quad m_H^{N=2} \; = \; m_A
\label{N=2treemasses}
\end{equation}
at the tree level. Moreover, at the tree level the $N=2$ Higgs sector is `aligned', so that the lighter
neutral Higgs boson has exactly the same couplings as in the Standard Model~\cite{EQS}.

\begin{figure}[ht]
\begin{center}
\includegraphics[width=7.5cm]{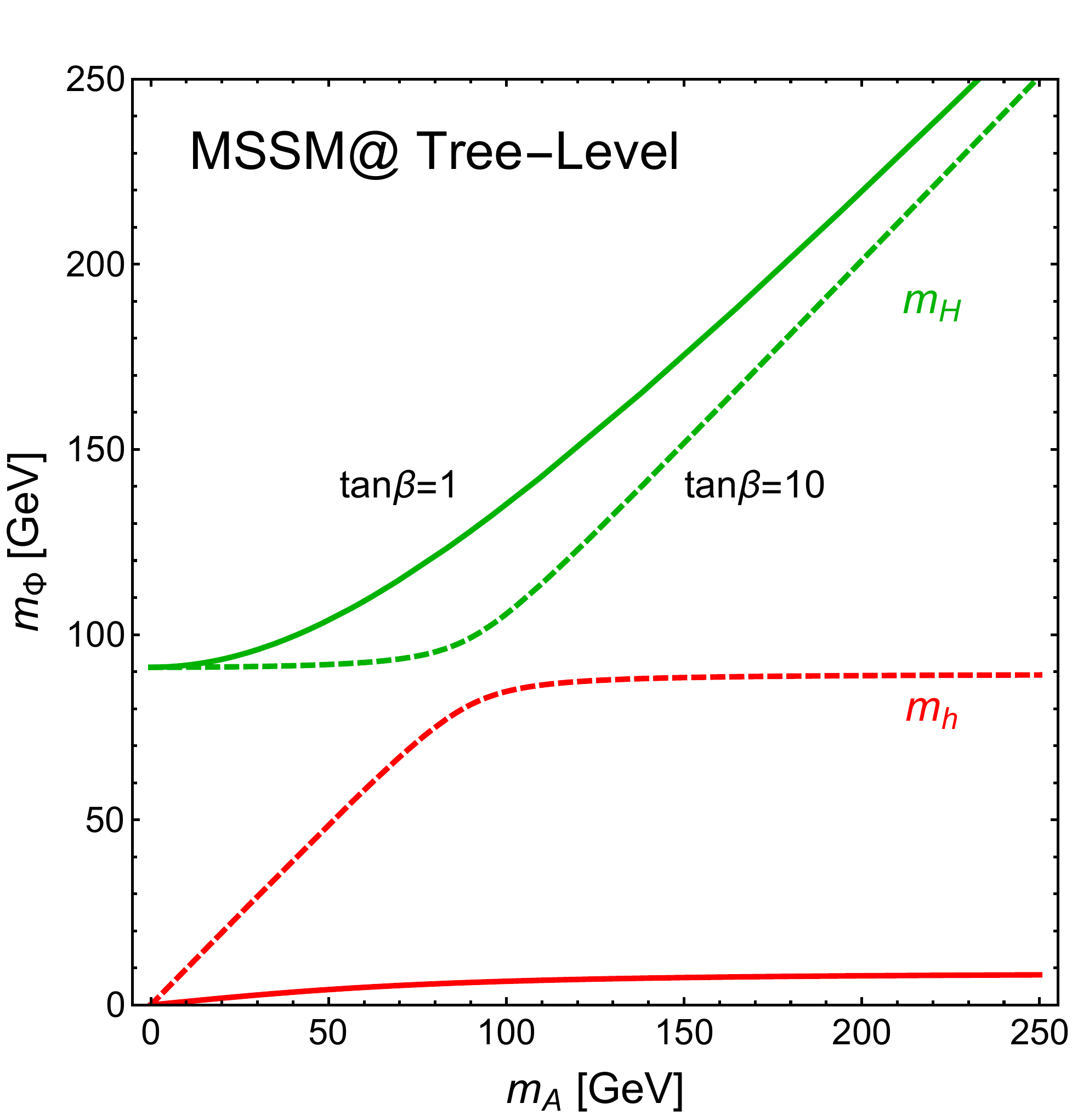}
\includegraphics[width=7.5cm]{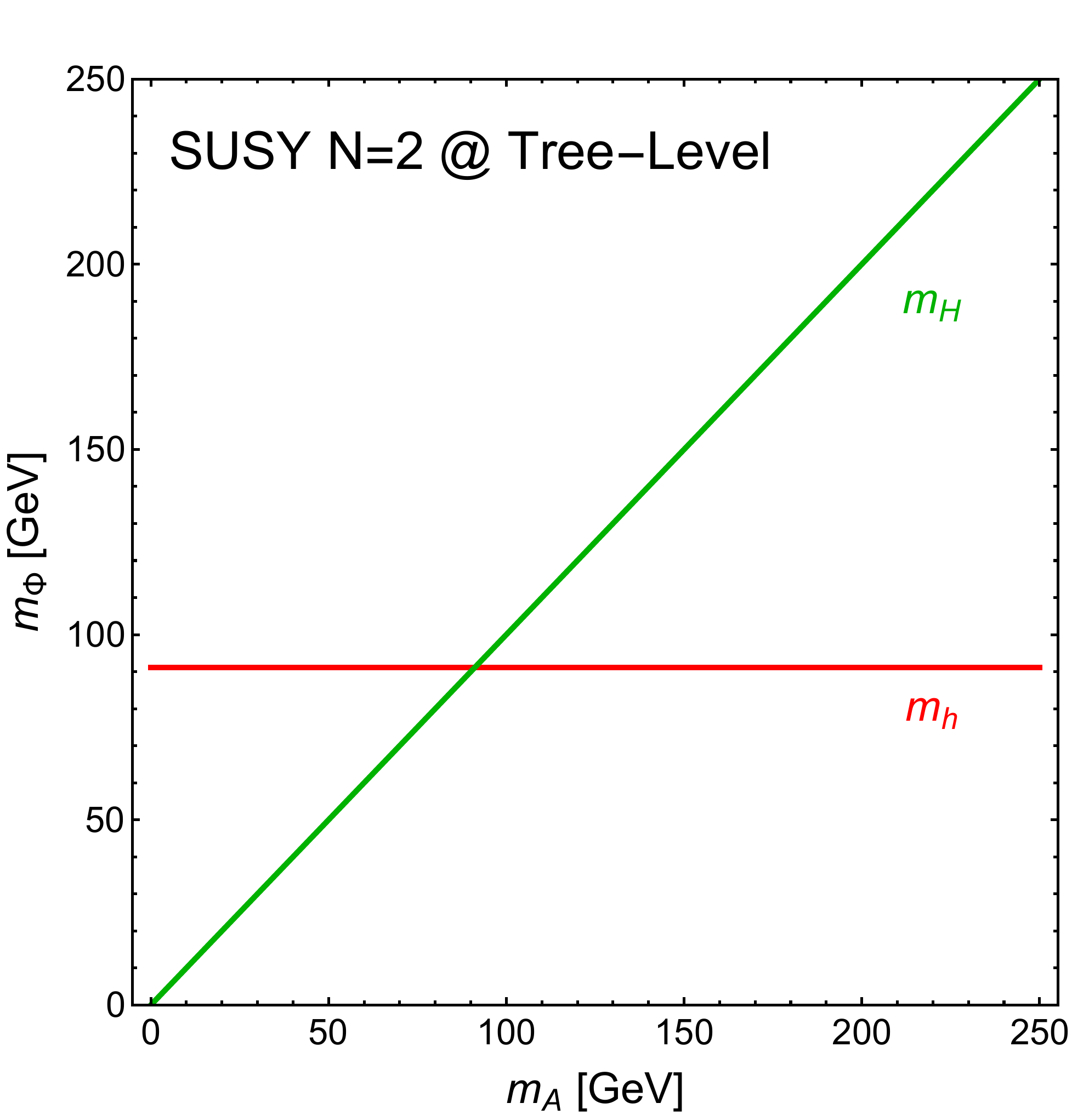}
\caption{\it A comparison between the tree-level values of Higgs masses in the MSSM
(left panel) and in the $N=2$ supersymmetric model (right panel)~\protect\cite{EQS}.}
\label{fig:treemasses}
\end{center}
\end{figure}

The dominant one-loop corrections $\varepsilon$ to (\ref{massmatrixN=1}) and (\ref{massmatrixN=2})
are those due to top quarks and stop squarks, which appear in their [22] entries. Let us
assume that they are such as to give the measured Higgs boson mass $m_h =125$~GeV.
In the $N = 1$ case the required loop correction is~\cite{hMSSM}
\begin{eqnarray}
\varepsilon_{N=1}= \Delta {\cal M}^{2,N=1}_{22}= \frac{m_{h}^2(m_{A}^2  + m_{Z}^2 -m_{h}^2) - m_{A}^2 m_{Z}^2 
\cos^2 2\beta } { m_{Z}^2 \cos^{2}{\beta}  +m_{A}^2 \sin^{2}{\beta} -m_{h}^2} \, ,
\label{dM22MSSM}
\end{eqnarray}
whereas in the $N = 2$ case it is~\cite{EQS}
\begin{eqnarray}
\varepsilon_{N=2}= \Delta {\cal M}^{2,N=2}_{22}=  \frac{2 (m_{A}^2-m_{h}^2) (m_{h}^2-m_{Z}^2)}{\cos 2 \beta \left(m_{Z}^2-m_{A}^2\right)+m_{A}^2-2 m_{h}^2+m_{Z}^2} \, .
\label{dM22h2MSSM}
\end{eqnarray}
Fig.~\ref{fig:compare} compares (left panel) the required value of $m_H$ for $\tan \beta = 1$ in the two cases,
(middle panel) the value of $m_H - m_A$ as a function of $m_A$ for $\tan \beta = 3$, 
and (right panel) the value of $m_H - m_A$ as a function of $\tan \beta$ for $m_A = 300$~GeV~\cite{EQS}.
We see that {\it $m_H$ can be lighter in the $N= 2$ case than when $N = 1$}, and that $m_H - m_A$
is also smaller, in general. Another effect of doubling up on supersymmetry in the Higgs sector is
seen in Fig.~\ref{fig:compareMSUSY}, where we compare the supersymmetry-breaking mass scale $M_{SUSY}$
that is needed in the $N=1$ and $N=2$ cases for different values of the squark mass miixing parameter $X_t$~\cite{EQS}.
We see a consistent pattern that {\it smaller values of $M_{SUSY}$ are required when $N=2$ than when
$N=1$} for any fixed values of $m_A$ and $\tan \beta$.

\begin{figure}[ht]
\begin{center}
\includegraphics[width=5.2cm]{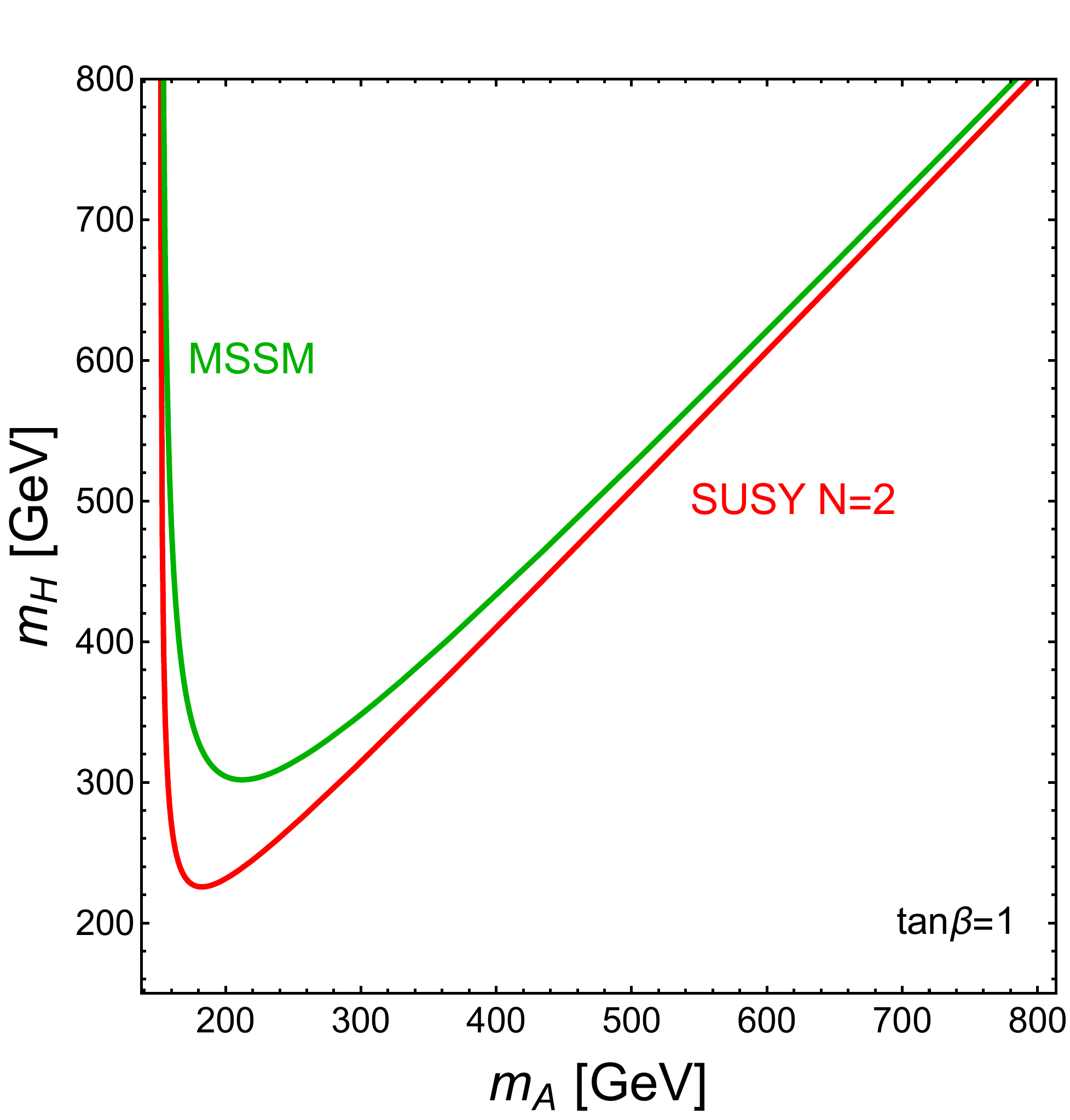}
\includegraphics[width=5.2cm]{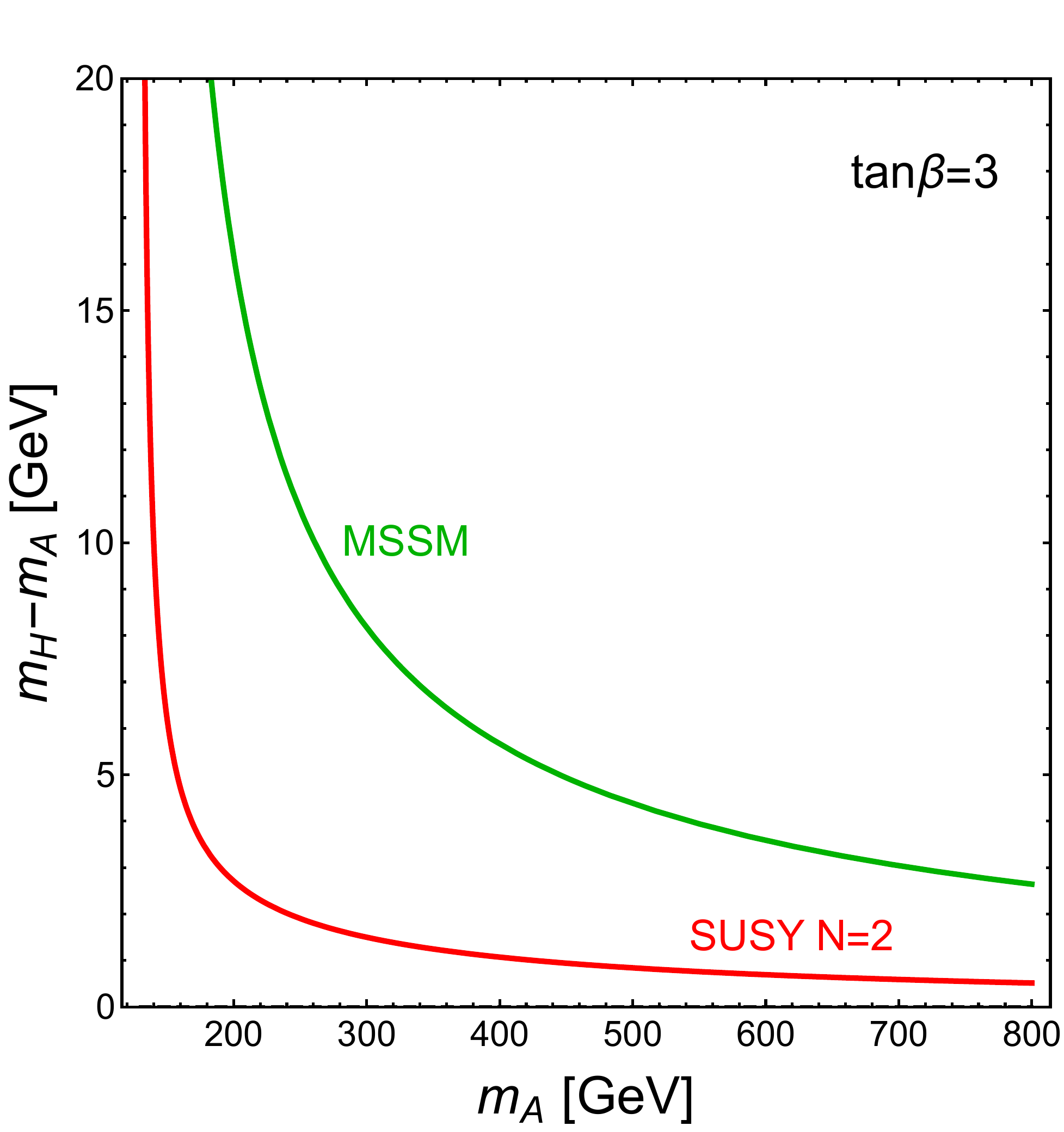}
\includegraphics[width=5.2cm]{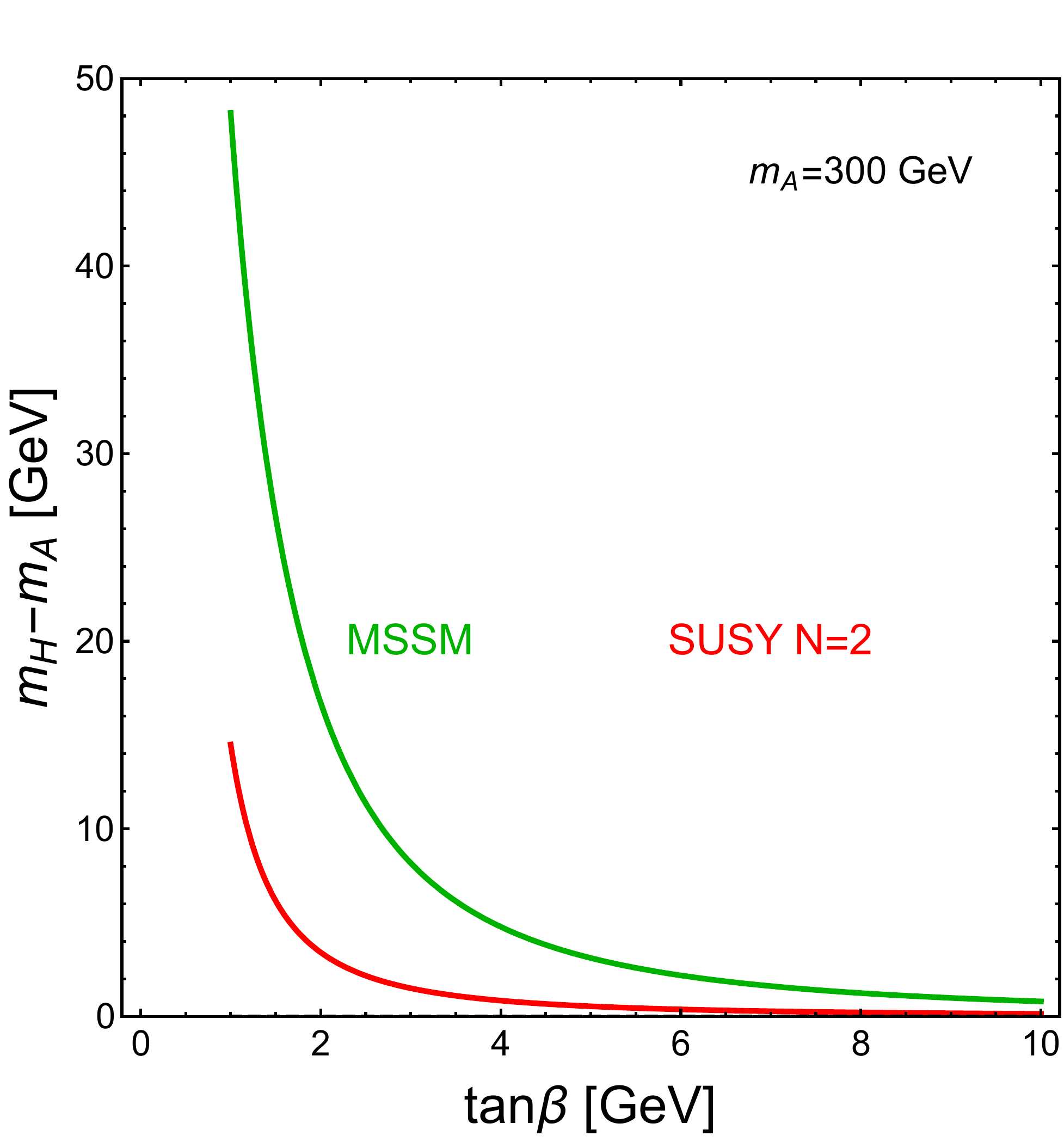}
\caption{\it Left panel: The value of $m_H$ required to obtain $m_h = 125$~GeV
via one-loop radiative corrections for $\tan \beta = 1$ in the MSSM and $N=2$ supersymmetry~\protect\cite{EQS}.
Middle panel: The value of $m_H - m_A$ as a function of $m_A$ for $\tan \beta = 3$. 
Right panel: The value of $m_H - m_A$ as a function of $\tan \beta$ for $m_A = 300$~GeV.}
\label{fig:compare}
\end{center}
\end{figure}

\begin{figure}[ht]
\begin{center}
\includegraphics[width=7cm]{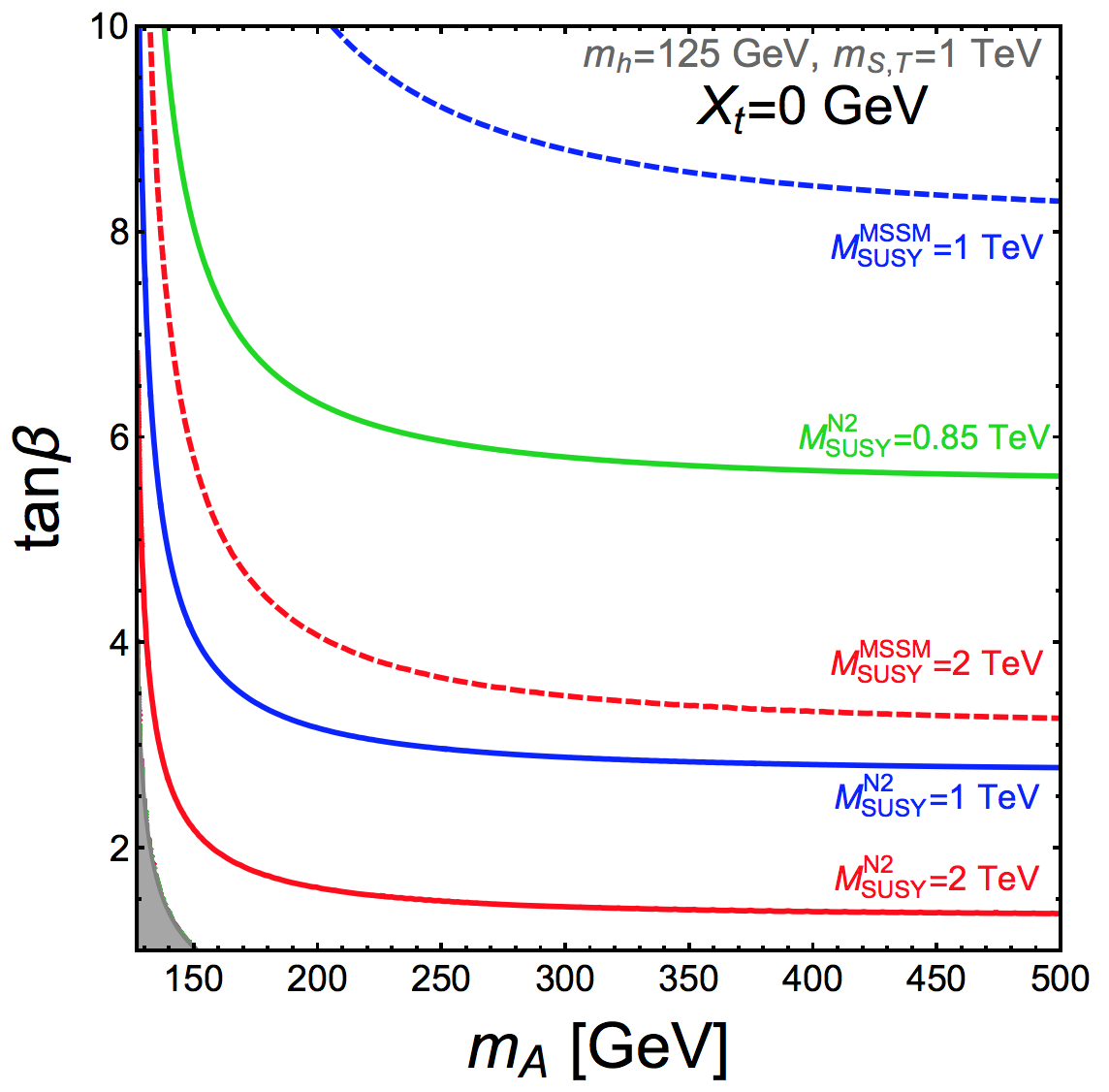}
\includegraphics[width=7cm]{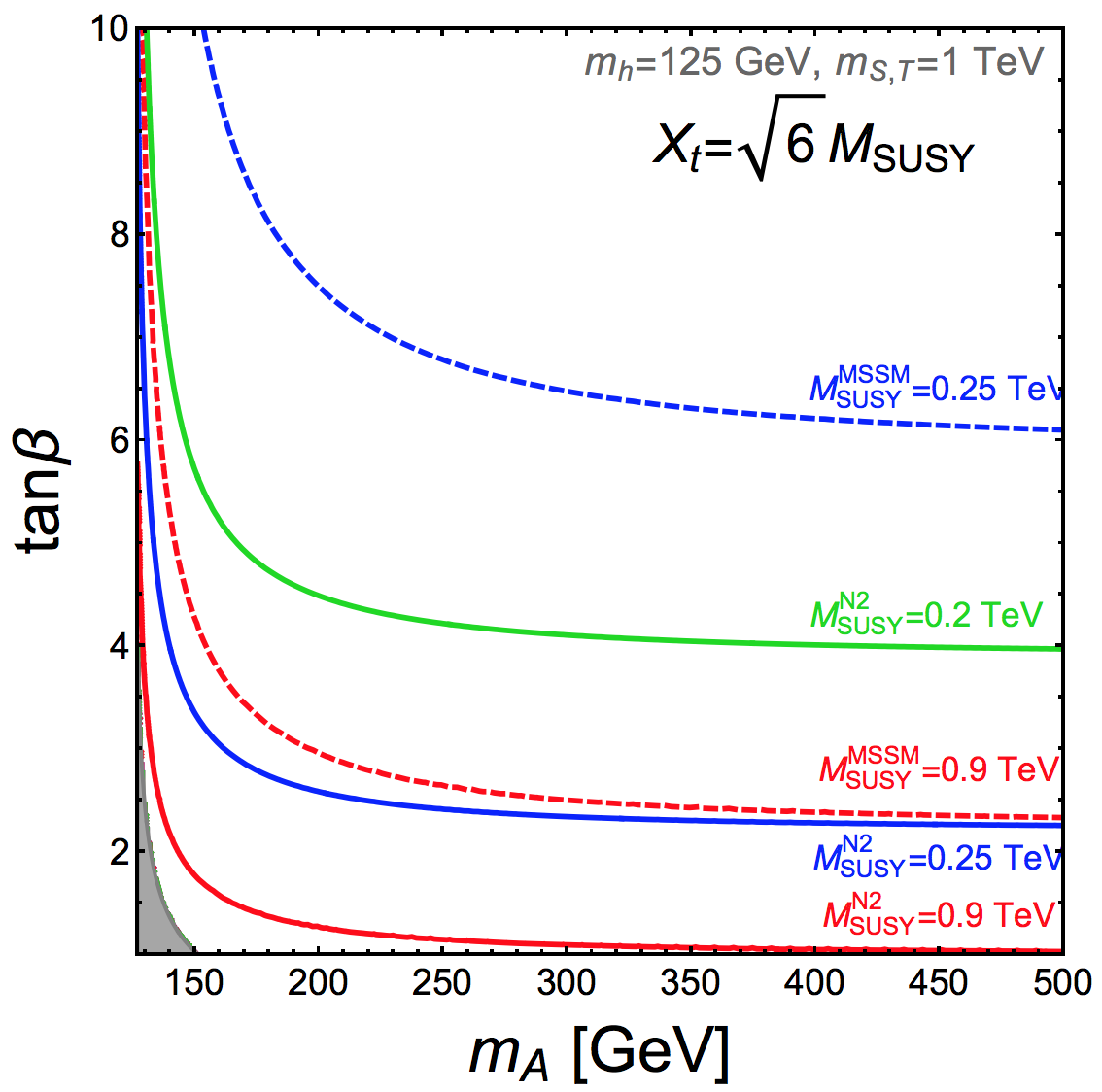}
\caption{\it Contours of the supersymmetry-breaking mass scale
$M_{SUSY}$ that are required as functions of $m_A$ and $\tan \beta$ to yield $m_h = 125$~GeV
in the MSSM scenario (dotted lines) and the $N=2$ scenario (full lines)~\protect\cite{EQS}.
The left panel is for $X_t = 0$, and the right panel is for
the maximal-mixing scenario with $X_t = \sqrt{6} M_{SUSY}$.}
\label{fig:compareMSUSY}
\end{center}
\end{figure}

On the other hand, the sensitivities of the LHC searches are also different, as seen in the
$(m_A, \tan \beta)$ plane in Fig.~\ref{fig:LHCHA}~\cite{EQS}. The upper part of this plane (shaded grey)
is excluded by direct LHC searches for $H, A \to \tau^+ \tau^-$, which have similar sensitivities
in the $N = 1$ and $N = 2$ cases. The red (green) curves show the ranges of $m_A$ that
are excluded indirectly by the LHC. We see that $m_A \gtrsim 200$~GeV is allowed in the
$N = 2$ case, whereas $m_A \gtrsim 350$~GeV is required when $N = 1$. The bottom line
is that {\it both supersymmetry and supersymmetric Higgs bosons may be closer
in an $N=2$ supersymmetric model than has been suggested by the $N = 1$ MSSM.}

\begin{figure}[ht]
\begin{center}
\includegraphics[width=9cm]{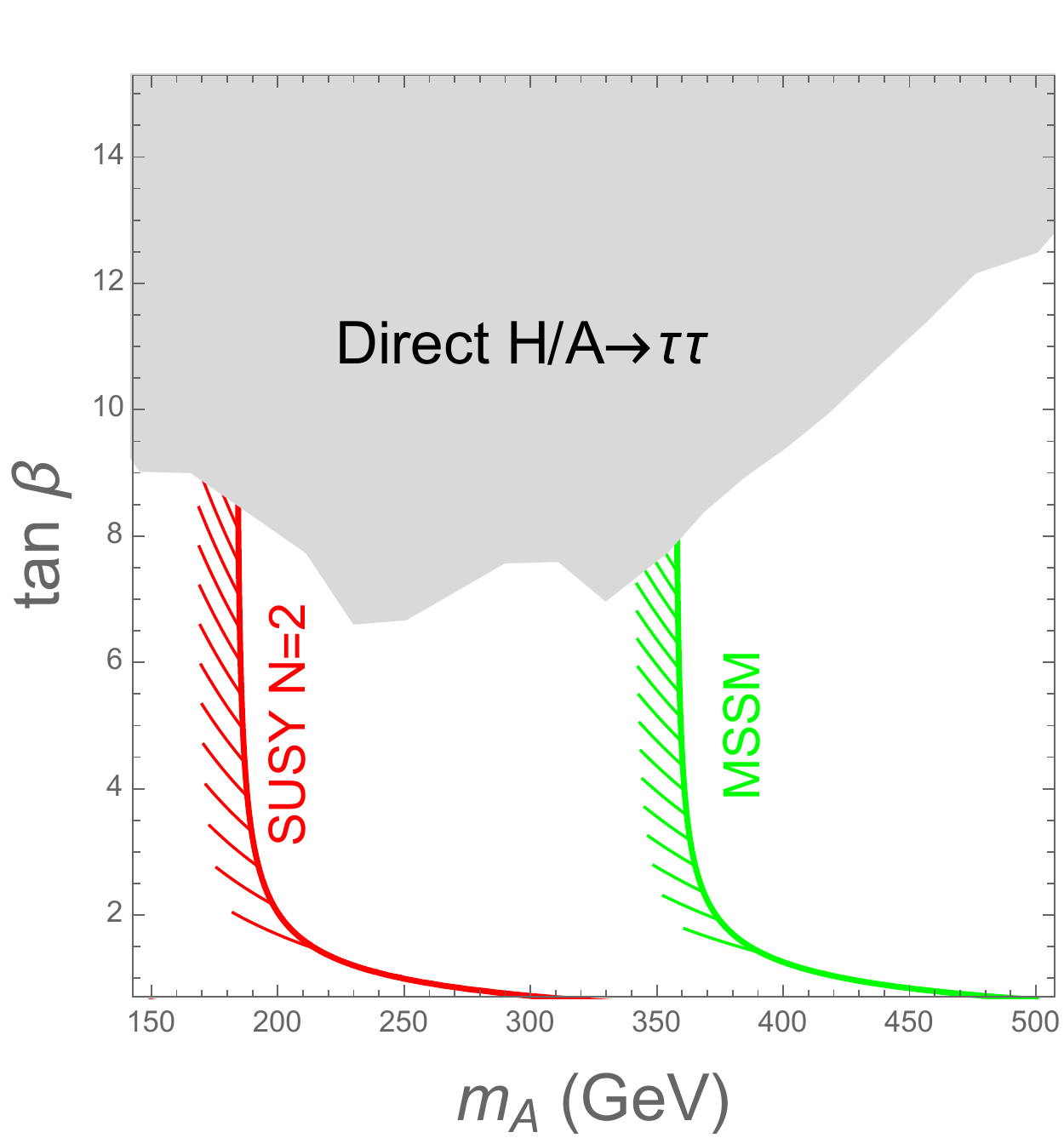}
\caption{\it The direct exclusion from searches for heavy scalars in the $H/A \to \tau \tau$ final state
(grey shading), and the indirect
bounds from measurements of Higgs couplings to fermions and massive bosons at Run~1 of the LHC
in the MSSM (green) and the $N=2$ model (red)~\protect\cite{EQS}.}
\label{fig:LHCHA}
\end{center}
\end{figure}

\subsection{What next: A Higgs factory?}

Now that a (the?) Higgs boson has been discovered, there is naturally a lot of interest
in studying it in detail. The LHC has considerable potential in this respect, with a target
of eventually accumulating 3000/fb of data with the HL-LHC that has now been approved
by the CERN Council~\cite{HL-LHC}. Ideas for future Higgs factories should take this into account,
and should be able to demonstrate how much better they can measure the Higgs boson,
as well as look for other possible new physics. Two proposals for linear $e^+ e^-$ colliders
are on the market: the ILC that aims initially at a centre-of-mass energy of 500 GeV with a planned upgrade to 1~TeV~\cite{ILC},
and CLIC that aims at centre-of-mass energies between 350~GeV and 3~TeV~\cite{CLIC}, as seen in Fig.~\ref{fig:e+e-}. 
Designs for circular $e^+ e^-$ colliders, CEPC in China~\cite{CEPC} and FCC-ee near CERN~\cite{FCC-ee}, are now also being discussed. 
As also seen in Fig.~\ref{fig:e+e-}, these are more limited in
centre-of-mass energy, but have the potential for higher luminosities for probing the
Higgs boson via the $e^+ e^- \to Z H$ process, and for precision electroweak studies
at the $Z$ peak and the $W^+ W^-$ threshold.

\begin{figure}[ht]
\begin{center}
\includegraphics[width=12cm]{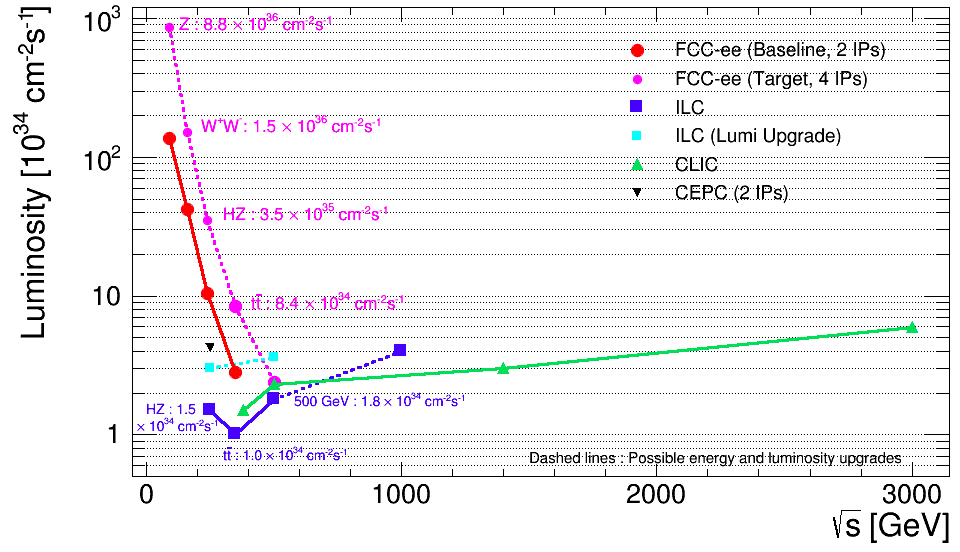}
\caption{\it The design luminosities at various centre-of-mass energies of projects
for future high-energy $e^+e^-$ colliders~\protect\cite{ILC,CLIC,CEPC,FCC-ee}.}
\label{fig:e+e-}
\end{center}
\end{figure}

The capabilities of the ILC~\cite{ILC} and FCC-ee~\cite{EYunpub} for Higgs coupling measurements are shown in 
the left and right panels of Fig.~\ref{fig:ILCFCC-ee}, respectively. The capabilities of
the LHC, HL-LHC, ILC and FCC-ee to probe the $H \gamma \gamma, H ZZ, HWW$
and $Hgg$ couplings are shown in Fig.~\ref{fig:SUSY}, and compared with the
deviations form the Standard Model that are expected in different supersymmetric models
whose parameters were chosen to be consistent with LHC data~\cite{Interplay}.
As mentioned previously, these typically predict very small deviations from the Standard Model
that will be very difficult to distinguish experimentally. Moreover, the current 
theoretical uncertainties in these predictions, indicated by the green bars, are large
compared with the prospective experimental accuracies at the FCC-ee, in particular.
More theoretical work will be needed!

\begin{figure}[ht]
\begin{center}
\includegraphics[width=6.8cm]{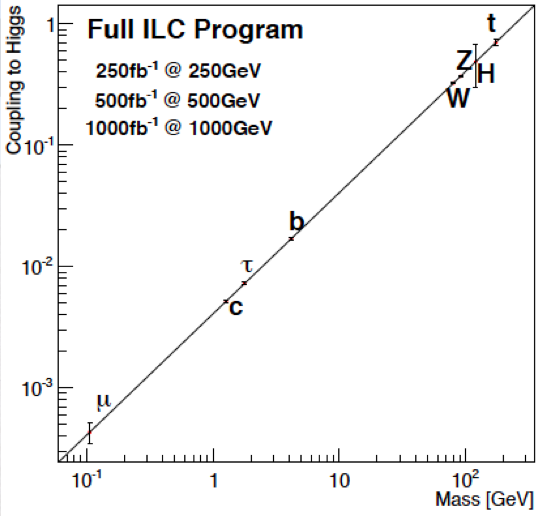}
\includegraphics[width=8.2cm]{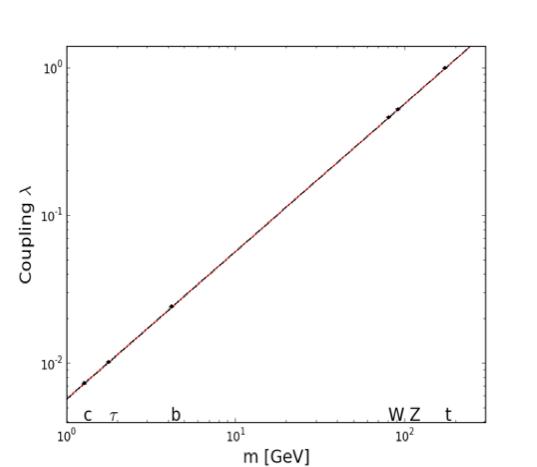}
\caption{\it Left panel: Prospective measurements of Higgs couplings at the ILC~\protect\cite{ILC}.
Right panel: Prospective measurements of Higgs couplings at FCC-ee~\protect\cite{EYunpub}.}
\label{fig:ILCFCC-ee}
\end{center}
\end{figure}

\begin{figure}[ht]
\begin{center}
\includegraphics[width=7cm]{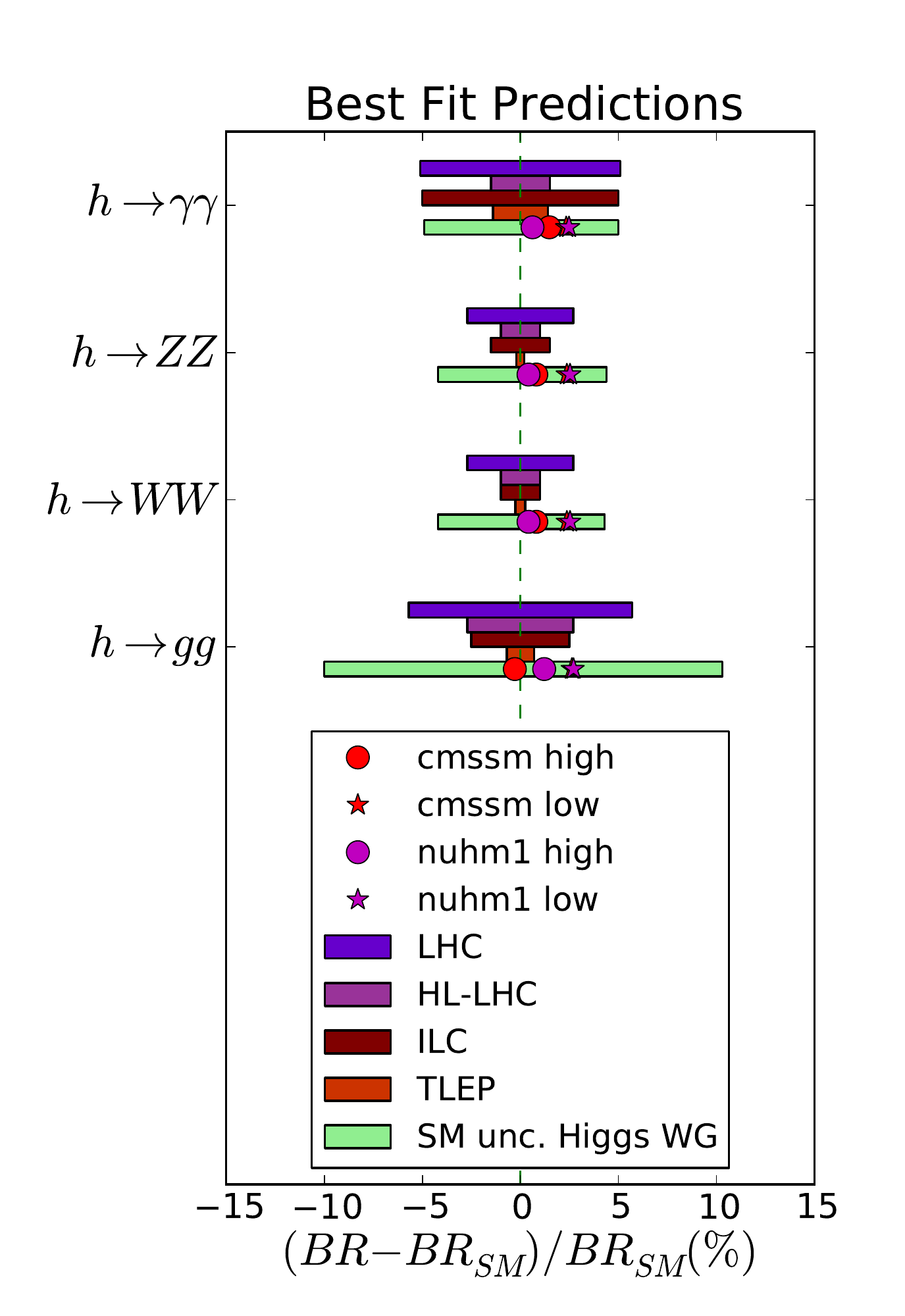}
\caption{\it Comparison between prospective
measurements of Higgs branching ratios at future colliders,  low- and high-mass CMSSM and
NUHM1 predictions (red and purple symbols) and the current uncertainties
within the Standard Model (turquoise bars)~\protect\cite{Interplay}.}
\label{fig:SUSY}
\end{center}
\end{figure}

As discussed earlier,
a favoured approach is to use future Higgs, precision electroweak and TGC measurements
to constrain the coefficients of possible dimension-6 operators constructed out of
Standard Model fields. Fig.~\ref{fig:LHCFCC-ee} compares the LHC constraints
(left panel) and the prospective FCC-ee sensitivities (right panel) in global fits to the scale $\Lambda$ in
dimension-6 operator coefficients~\cite{EYFCC-ee}. Fig.~\ref{fig:ILCFCC-ee6} compares the
sensitivities of the ILC and FCC-ee in fits combining Higgs and precision electroweak data
(left panel), and combining Higgs and TGC data (right panel)~\cite{EYFCC-ee}. In the left panel, the shadings 
compare results obtained with and without estimates of the theoretical uncertainties in the 
precision electroweak observables: we see again the importance of minimizing these. In the
right panel, the effects of including TGC measurements at the ILC are also indicated
by shading.

\begin{figure}[ht]
\begin{center}
\includegraphics[width=7.7cm]{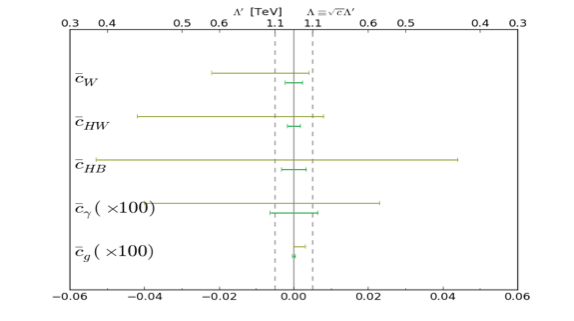}
\includegraphics[width=7.7cm]{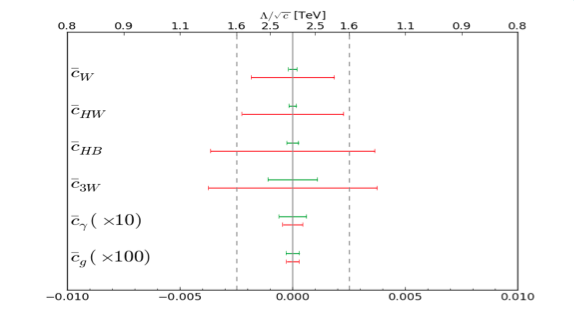}
\caption{\it Left panel: Constraints on dimension-6 operator coefficients from measurements at the LHC.
Right panel: Prospective corresponding constraints from measurements at FCC-ee~\protect\cite{EYFCC-ee}.}
\label{fig:LHCFCC-ee}
\end{center}
\end{figure}

\begin{figure}[ht]
\begin{center}
\includegraphics[width=7.5cm]{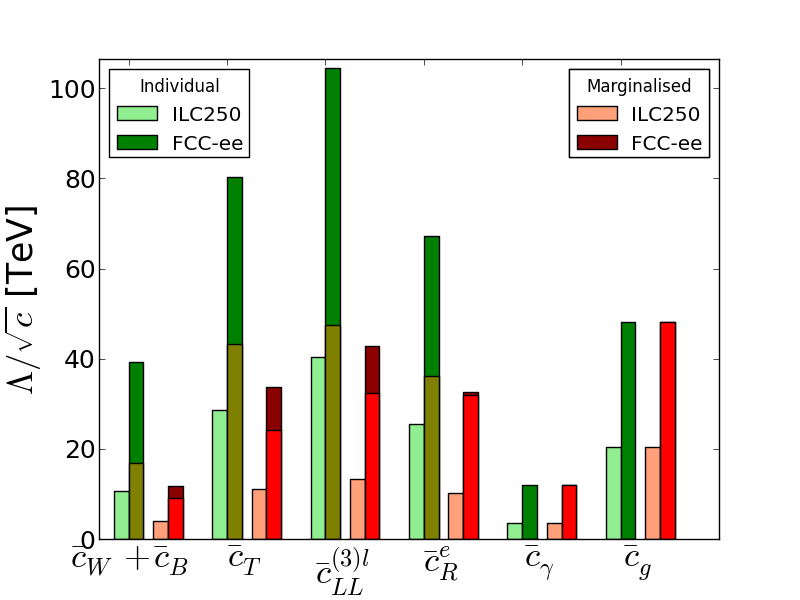}
\includegraphics[width=7.5cm]{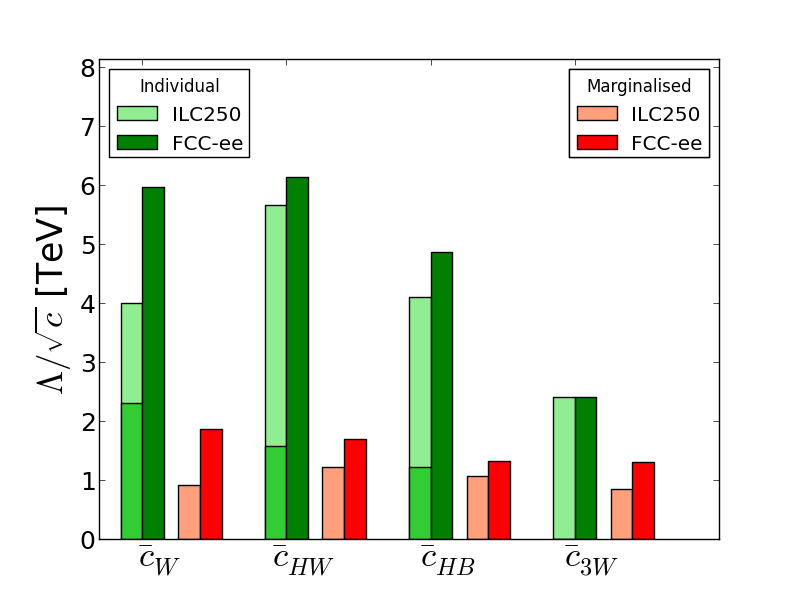}
\caption{\it Summary of the reaches for the dimension-6 operator coefficients, in individual
fits (green) and when marginalised in a global fit including all operators (red), 
from projected ILC250 (lighter shades) and FCC-ee (darker shades) precision measurements~\protect\cite{EYFCC-ee}. 
The left plot shows the operators that are most strongly constrained by electroweak precision
observables and Higgs physics, and the different shades of dark green and dark red
illustrate the effects of theoretical uncertainties at FCC-ee. The right plot is constrained
primarily by Higgs physics and TGCs, and the different shades of light green demonstrate the 
improved sensitivity when TGCs are included in the ILC250 analysis.}
\label{fig:ILCFCC-ee6}
\end{center}
\end{figure}

By virtue of its higher centre-of-mass energy, the CLIC proposal for an $e^+ e^-$ collider 
has particular advantages in looking for the effects of dimension-6 operators,
since the effects of their interferences with Standard Model amplitudes typically
increase $\propto E^2$~\cite{ERSY}. The following are some examples of the sensitivities to dimension-6
operator coefficients at 350~GeV and 3~TeV for associated $H Z$ production:
\begin{align}
\left. \frac{\Delta\sigma(HZ)}{\sigma(HZ)}\right|_\text{350 GeV} &= 16\bar{c}_{HW} + 4.7\bar{c}_{HB} + 35\bar{c}_W + 11\bar{c}_B - \bar{c}_H + 5.5\bar{c}_\gamma \, , \nonumber \\
\left. \frac{\Delta\sigma(HZ)}{\sigma(HZ)}\right|_\text{3 TeV} &= 2130\bar{c}_{HW} + 637\bar{c}_{HB} + 2150\bar{c}_W + 193\bar{c}_B - \bar{c}_H + 7.4\bar{c}_\gamma \, ,
\label{eq:HZcrosssections}
\end{align}
and for $e^+e^- \to W^+W^-$ production : 
\begin{align}
\left. \frac{\Delta\sigma(W^+W^-)}{\sigma(W^+W^-)}\right|_\text{350 GeV} &= 0.63\bar{c}_{HW} + 0.31\bar{c}_{HB}  + 4.6\bar{c}_W - 0.43\bar{c}_{3W} \, , \nonumber \\
\left. \frac{\Delta\sigma(W^+W^-)}{\sigma(W^+W^-)}\right|_\text{3 TeV} &= 13\bar{c}_{HW} + 7.8\bar{c}_{HB} + 17\bar{c}_W - 4.4\bar{c}_{3W} \, .
\label{eq:WWcrosssections}
\end{align}
The sensitivities to most of the operator coefficients indeed increase substantially with the
centre-of-mass energy, for both the associated $H + Z$ and $W^+ W^-$ cross-sections,
confirming the expected competitive advantage of the high energies attainable with CLIC.
Fig.~\ref{fig:CLIC6} shows the increase in sensitivity of CLIC 
operating at 3~TeV compared with 350~GeV or 1.4~TeV for a number of dimension-6
operator coefficients~\cite{ERSY}. The green bars are for fits including individual operators, and the
red bars are for global fits with the coefficients marginalized.

\begin{figure}[ht]
\begin{center}
\includegraphics[width=7.5cm]{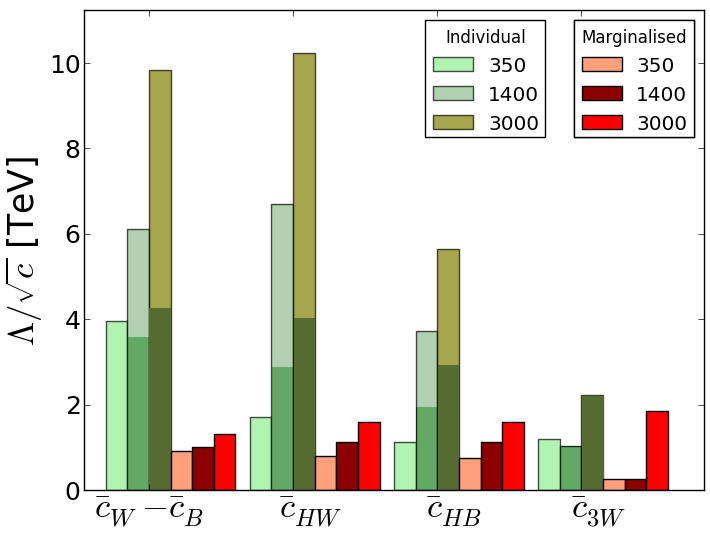}
\includegraphics[width=7.5cm]{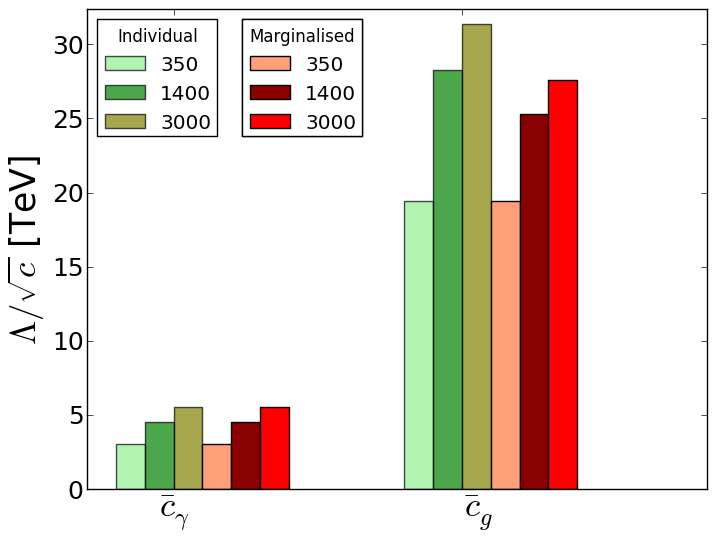}
\caption{\it The estimated sensitivities of CLIC measurements at 
centre-of-mass energies of 350~GeV, 1.4~TeV and 3.0~TeV to the scales
of various (combinations of) dimension-6 operator coefficients~\protect\cite{ERSY}.
The results of individual (marginalised) fits are shown as green (red) bars.
The lighter (darker) green bars in the left panel
include (omit) the prospective $HZ$ Higgsstrahlung constraint.}
\label{fig:CLIC6}
\end{center}
\end{figure}

Studies of the sensitivities to Higgs properties of a 100-TeV pp collider such as FCC-hh
are still at an early stage. However, as seen in Fig.~\ref{fig:FCC-hh}, the Higgs production
cross sections will be much larger than at the LHC, and there will be
extensive opportunities for kinematical measurements as well as overall production rates~\cite{FCC-hh}.
Moreover, such a machine might provide the first opportunity to measure directly the triple-Higgs
coupling with respectable accuracy, since it contributes to the $HH$ cross section that increases
by almost two orders of magnitude compared to the LHC, as seen in Fig.~\ref{fig:FCC-hh} (grey line).

\begin{figure}[ht]
\begin{center}
\includegraphics[width=10cm]{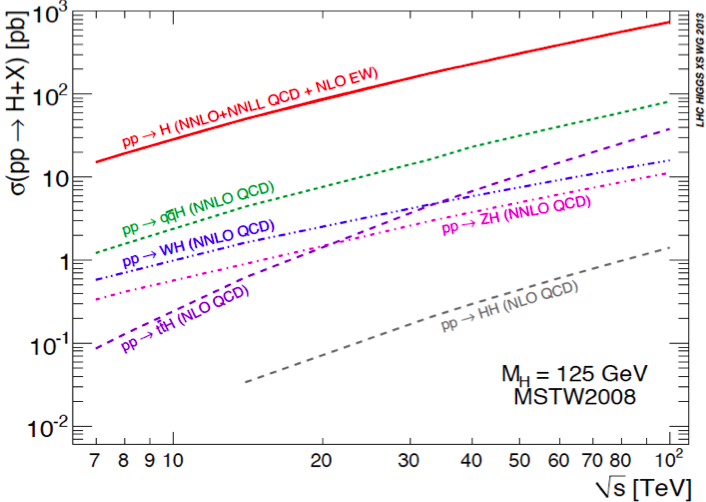}
\caption{\it The most important cross sections for Higgs productions in
pp collisions, as functions of the centre-of-mass energy up to 100 TeV~\protect\cite{FCC-hh}.}
\label{fig:FCC-hh}
\end{center}
\end{figure}

\subsection{Final remarks}

``Beyond any reasonable doubt", the LHC has discovered a (possibly the) Higgs boson~\cite{NP}.
Whilst being a tremendous success for theoretical physics, it also represents a
tremendous challenge. Even in the minimal elementary Higgs model, both the terms
in the Higgs potential present problems. Does the quartic term turn negative at high scales,
implying vacuum instability? How come the quadratic term is so small compared to
plausible fundamental mass scales in physics such as the Planck mass?
The LHC may yet discover new physics beyond the Standard Model during Run~2.
If it does, the global priority for high-energy physics will surely be to study it.
If it does not discover new physics at the TeV scale, it will be natural to focus future
accelerator experiments on the Higgs boson. Either way, in my personal opinion
future circular colliders may offer the best experimental prospects, being able to
probe the 10~TeV scale indirectly via high-precision low-energy experiments, and
directly via the production of new heavy particles.

\section*{Acknowledgements}

The author thanks the UK STFC for support via the research
grant ST/J002798/1.

\end{document}